\documentclass[manuscript]{emulateapj}

\usepackage{graphicx}
\usepackage{tabulary}
\usepackage{epsfig}
\usepackage{amssymb}
\usepackage{multirow}
\usepackage{appendix}
\usepackage{natbib}
\newcommand{\JGG}{J.Geomag.Geoelectr.}
\newcommand{\icarus}{Icarus\ }  
\newcommand{\beq}[1]{\begin{equation}\label{#1}}
\newcommand{\eeq}{\end{equation}}
\newcommand{\ME}{M_{\oplus}}
\newcommand{\MSUN}{M_{\odot}}
\newcommand{\Ms}{M_{\star}}

\renewcommand{\bar}{\overline}
\newcommand{\rom}[1]{{\rm #1}}

\shorttitle{Evolution of magnetic protection in habitable terrestrial
  planets}

\shortauthors{Zuluaga, Cuartas-Restrepo, Hoyos}

\begin{document}

%%%%%%%%%%%%%%%%%%%%%%%%%%%%%%%%%%%%%%%%%%%%%%%%%%%%%%%%%%%%%%%%%%%%%%%%%%%%%%%%%
%FRONT MATTER
%%%%%%%%%%%%%%%%%%%%%%%%%%%%%%%%%%%%%%%%%%%%%%%%%%%%%%%%%%%%%%%%%%%%%%%%%%%%%%%%%

\title{Evolution of magnetic protection in potentially habitable
  terrestrial planets}

\author{Jorge I. Zuluaga\altaffilmark{1}, Pablo
  A. Cuartas-Restrepo\altaffilmark{2}} \affil{Instituto de F\'isica -
  FCEN, Universidad de Antioquia,\\ Calle 67 No. 53-108, Medell\'in,
  Colombia}

\and

\author{Jaime H. Hoyos\altaffilmark{3}}
\affil{Departamento de Ciencias B\'asicas, Universidad de
        Medell\'in,\\ Carrera 87 No. 30-65, Medell\'in, Colombia}

\altaffiltext{1}{jzuluaga@fisica.udea.edu.co}
\altaffiltext{2}{p.cuartas@fisica.udea.edu.co}
\altaffiltext{3}{jhhoyos@udem.edu.co}

%%%%%%%%%%%%%%%%%%%%%%%%%%%%%%%%%%%%%%%%%%%%%%%%%%%%%%%%%%%%%%%%%%%%%%%%%%%%%%%%%
%ABSTRACT
%%%%%%%%%%%%%%%%%%%%%%%%%%%%%%%%%%%%%%%%%%%%%%%%%%%%%%%%%%%%%%%%%%%%%%%%%%%%%%%%%

\begin{abstract}
We present here a comprehensive model for the evolution of the
magnetic properties of habitable terrestrial planets (Earth-like
planets, $M\sim 1\,M_\oplus$ and super-Earths $M\sim\,1-10\,M_\oplus$)
and their effects on the long-term protection against the atmospheric
erosive action of stellar wind.  Using up-to-date thermal evolution
models and dynamo scaling laws we predict the evolution of the
planetary dipole moment as a function of planetary mass and rotation
rate.  Combining these results with models for the evolution of the
stellar wind, stellar XUV fluxes and exosphere properties of highly
irradiated planets, we determine the properties of the planetary
magnetosphere and the expected scale height of the atmosphere that
together determine the level of thermal and non-thermal atmospheric
mass losses. We have used this model to evaluate the early magnetic
protection of the Earth and the already discovered potentially
habitable super-Earths GJ 667Cc, Gl 581d and HD 85512b.  We confirm
that Earth-like planets, even under the highest attainable
dynamo-generated magnetic field strengths, will lose a significant
fraction of their atmospheres or their content of critical volatiles
(e.g. H$_2$O) if they are tidally locked in the HZ of dM stars.
Planets in this mass-range with N/O-rich atmospheres, even under the
best conditions of magnetic protection, will probably lose their
atmospheres or their water content if they are in habitable zones
closer than $\sim$ 0.8 AU ($M_\star\lesssim 0.7-0.9 M_\odot$).
Super-Earths $M_p\gtrsim 3 M_\oplus$ seem to have better chances of
preserving their atmospheres even if they are tidally locked around dM
stars. Under similar conditions of thermal and magnetic field
evolution there seems to exist a planetary mass-dependent inner limit
inside the HZ itself, below which large atmospheric mass-losses in
super-Earths are expected.  With the nominal value of the physical
parameters in our conservative model this limit is, for example,
$\sim$0.1 AU for $4 \ME$ and $\sim$0.04 AU for $8 \ME$.  Under these
conditions we predict that the atmosphere of GJ 667Cc has probably
already been obliterated and it is presently uninhabitable.  On the
other hand, our model predicts that the atmospheres of Gl 581d and HD
85512b would be well protected by dynamo-generated magnetic fields
even under the worst expected conditions of stellar aggression.
\vspace{0.3cm}
\end{abstract}
%REVISED22

%%%%%%%%%%%%%%%%%%%%%%%%%%%%%%%%%%%%%%%%%%%%%%%%%%%%%%%%%%%%%%%%%%%%%%%%%%%%%%%%%
%KEY WORDS
%%%%%%%%%%%%%%%%%%%%%%%%%%%%%%%%%%%%%%%%%%%%%%%%%%%%%%%%%%%%%%%%%%%%%%%%%%%%%%%%%


%%%%%%%%%%%%%%%%%%%%%%%%%%%%%%%%%%%%%%%%%%%%%%%%%%%%%%%%%%%%%%%%%%%%%%%%%%%%%%%%%
%PAPER CONTENT
%%%%%%%%%%%%%%%%%%%%%%%%%%%%%%%%%%%%%%%%%%%%%%%%%%%%%%%%%%%%%%%%%%%%%%%%%%%%%%%%%

%%%%%%%%%%%%%%%%%%%%%%%%%%%%%%%%%%%%%%%%%%%%%%%%%%%%%%%%%%%%%%%%%%%%%%%%%
\section{Introduction}
\label{sec:introduction}
%%%%%%%%%%%%%%%%%%%%%%%%%%%%%%%%%%%%%%%%%%%%%%%%%%%%%%%%%%%%%%%%%%%%%%%%%

The discovery of extrasolar habitable planets is one of the most
ambitious challenges in the exoplanetary research.  At the time of
writing, there are almost 760 confirmed exoplanets \footnote{For
  updates, please refer to {http://exoplanet.eu}} including 55
classified as Earth-like planets (EPs, $M\sim 1\,M_\oplus$) and
super-Earths (SEs, $M\sim\,1-10\,M_\oplus$, \citealt{Valencia06}).
Although the composition of most of these planets is unknown, many of
them should have similar compositions to the Earth which would make
them the first extrasolar terrestrial planets (TPs) discovered so far.
%REVISED2

Among these low mass planets already discovered there are three
confirmed SEs, GJ 667Cc, Gl 581d and HD 85512b
\citep{Selsis07a,Pepe11,Kaltenegger11} and tens of Kepler candidates
\citep{Borucki11, Batalha12} that are close or inside the habitable
zone (HZ) of their host stars
%\footnote{For updates a detailed analysis of the habiltability of
%  presently confirmed exoplanets please refer to the {\it Habitability
%    Laboratory} {http://phl.upr.edu}}.  
If we include the possibility
that extra-solar-system giant planets could harbour habitable
exomoons, the number of already discovered potentially habitable
planetary environments beyond the Solar System could be rised to
several tens \citep{Kaltenegger10a,Underwood03}.  Moreover, the case
for the existence of a plethora of other TPs and exomoons in the
Galaxy is rapidly gaining evidence
\citep{Borucki11,Catanzarite11,Kipping12} and the chances that a large
number of potentially habitable extrasolar bodies could be discovered
in the near future are encouraging.
%REVISED2

The question of which properties a planetary environment needs in
order to allow the appearance, evolution and diversification of life
has been extensively studied (for recent reviews see
\citealt{Lammer09} and \citealt{Kasting10}).  Two basic and
complimentary physical conditions must be fulfilled: the presence of
an atmosphere and the existence of surface liquid water
\citep{Kasting93}.  However, the fulfilment of these basic conditions
depends on many complex and diverse endogenous and exogenous factors
(for a comprehensive enumeration of these factors see
e.g. \citealt{Ward00} or \citealt{Lammer10})
%REVISED2

The existence and long-term stability of an intense planetary magnetic
field (PMF) is one of these additional factors (see
e.g. \citealt{Griebmeier10} and references therein).  It has been
shown that a strong enough PMF would protect the atmosphere of
potentially habitable planets, especially its valuable content of
water and other volatiles, against the erosive action of the stellar
wind \citep{Lammer03,Lammer07,Khodachenko07,Chaufray07}.  Planetary
magnetospheres would also act as shields against the potentially
harmful effects that the stellar and galactic cosmic rays (CR) could
produce in the life-forms evolving on its surface (see
e.g. \citealt{Griebmeier05}).  Even in the case that life could arise
and evolve on unmagnetized planets, the detection of atmospheric
biosignatures would be also affected by a higher flux of stellar and
galactic CR, especially if the planet is around very active M-dwarfs
(dM) \citep{Grenfell07,Segura10}.  In summary, a PMF would not only
protect the atmosphere of the planet and the life growing on its
surface, but also give us the possibility to confirm the habitability
of future discovered planets around dM stars.
%REVISED2

But find suitable conditions to have TPs with strong enough PMFs in
the HZ of the most abundant stars seems more problematic than
previously thought.  It has been recently predicted that most of the
TPs in our Galaxy could be found around dM stars \citep{Boss06,
  Mayor08, Scalo07, Rauer11}. Actually $\sim 20\%$ of the presently
confirmed super-Earths belong to planetary systems around stars of
this type, including Gl 581d one of the best candidate for
habitability presently known \citep{Selsis07a}.  Planets inside the HZ
of low mass stars ($\Ms \lesssim 0.6 \MSUN$) would be tidally locked
\citep{Joshi97,Heller11} a condition that poses serious limitations to
their potential habitability (see e.g. \citealt{Kite11} and references
therein).  Tidally locked planets inside the HZ of dMs will have
periods in the range of $5-100$ days, a condition has commonly been
associated with the almost complete lack of a protective magnetic
field \citep{Griebmeier04}.  However, the relation between rotation
and PMF properties, that is critical to assess the magnetic protection
of slowly rotating planets, is more complex than previously thought
\citep{Zuluaga12}.  In particular a detailed knowledge of the thermal
evolution of the planet is required to predict not only the intensity
but also the regime (dipolar or multipolar) of the PMF for a given
planetary mass and rotation rate.
%REVISED2

Several authors have extensively studied the protection that intrinsic
PMF would provide to extrasolar planets \citep{Griebmeier05,
  Khodachenko07, Lammer07, Griebmeier09, Griebmeier10}.  Independently
the effects of X-Ray and EUV radiation (XUV) on the thermal escape
processes in weakly magnetized planets have also been studied
\citep{Kulikov06,Lammer07,Lammer09,Tian08,Tian09,Sanz10}.  All these
works have however neglected the evolving nature of the PMF and have
systematically predicted field intensities using scaling-laws that
have been revised in the last couple of years (see
\citealt{Christensen10} and references therein).  More importantly the
role of rotation in determining the PMF properties that is critical in
assessing the case of tidally locked planets has been overlooked
\citep{Zuluaga12}.
%REVISED2

In this work we present a model of the evolution of the magnetic
protection of potentially habitable TPs around GKM main sequence
stars.  To achieve this goal we integrate in a single framework the
most recent thermal evolution models for this type of planets
\citep{Gaidos10,Tachinami11}, up-to-date dynamo scaling-laws
\citep{Christensen10,Zuluaga12}, models for the evolution of the
stellar wind and XUV luminosity of low mass-stars
\citep{Griebmeier10,Sanz11} and the most recent results describing the
expansion and hydrodynamical escape of volatiles from highly
irradiated atmospheres of low mass planets
\citep{Kulikov06,Tian08,Tian09}.  This is the first time that all
these pieces have been put together to produce a global picture of the
magnetic protection of potentially habitable TPs.
%REVISED2

But this is not only a model integration effort.  Several novel
features have been added to our {\it comprehensive model}: 1) we
include a new treatment of the role of rotation in determining the PMF
properties, eespecially important in assessing the magnetic protection
of tidally locked planets, 2) we propose a phenomenological formula to
estimate a {\it magnetic-constrained} thermal mass-loss rate from
atmospheres protected with a strong PMF, and 3) we address the
magnetic protection of already discovered habitable planets and
compare it with the case of an early magnetized Venus and the Earth in
its current state.
%REVISED2

This paper is organized as follows: section
\ref{sec:MagnetosphereModel} describes the model used here to
calculate the evolution of magnetosphere properties.  For that purpose
we use a set of analytical fits of recently published models for the
thermal evolution of TPs (section \ref{subsec:ThermalEvolution}) and
up-to-date power-based dynamo scaling laws (section
\ref{subsec:ScalingPMF}).  Several stellar key properties are required
to calculate the evolution of the magnetopshere (stellar wind
properties, HZ distances and tidally locking limits).  The way we
obtain these properties and their evolution are described in section
\ref{subsec:StellarProperties}.  Atmospheric expansion models and the
proposed phenomenological formula to model the atmospheric mass-loss
rate are described in section \ref{sec:AtmosphericEscape}.  Section
\ref{sec:Results} presents the results of applying this model to
hypothetical and already discovered habitable TPs.  A discussion about
the hypothesis on which the model relies and the possible sources of
uncertainties are presented in section \ref{sec:Discussion}.  Finally
several conclusions and the future prospects of this work are
summarized in section \ref{sec:Conclusions}.
%REVISED2

%%%%%%%%%%%%%%%%%%%%%%%%%%%%%%%%%%%%%%%%%%%%%%%%%%%%%%%%%%%%%%%%%%%%%%%%%
\section{A model for an evolving magnetosphere}
\label{sec:MagnetosphereModel}
%%%%%%%%%%%%%%%%%%%%%%%%%%%%%%%%%%%%%%%%%%%%%%%%%%%%%%%%%%%%%%%%%%%%%%%%%

The interaction between the PMF, the interplanetary magnetic field
(IMF) and the stellar wind creates a magnetic cavity around the planet
known as the magnetosphere. Magnetospheres are very complex systems
but its basic properties are continuous functions of two basic
variables \citep{Siscoe75b}: the planetary magnetic dipole moment
${\cal M}$ and the dynamical pressure of the stellar wind
$P_\rom{sw}$:
%REVISED2

\beq{eq:M}
{\cal M}=\frac{4\pi r^3 B_{p}^\rom{dip}(r)}{\sqrt{2}\mu_0},
\eeq

\beq{eq:Psw}
P_\rom{sw}=m n v_\rom{eff}^2+2 n k_B T.
\eeq

Here $B_{p}^\rom{dip}(r)$ is the dipolar component of the field as
measured at distance $r$ from the planet center. $m$ and $n$ are the
typical mass of a wind particle (mostly protons) and its number
density, respectively. $v_\rom{eff}=(v_\rom{sw}^2+v_{p}^2)^{1/2}$ is
the effective average velocity of the stellar wind as measured in the
reference frame of the planet whose orbital velocity is $v_p$.  And
$T$ is the local temperature of the plasma.  $\mu_0=4 \pi \times
10^{-7}\,\rm{H/m}$ and $k_B$ are the vacuum permeability and Boltzmann
constant respectively.
%REVISED2

There are three basic properties of planetary magnetospheres we are
interested in: 1) the maximum magnetopause field intensity $B_{mp}$, a
proxy to the flux of high energy particles into the magnetospheric
cavity, 2) the standoff or stagnation radius, $R_S$, a measure of the
size of the dayside magnetosphere, and 3) the area of the polar cap
$A_{pc}$ that measures the total area of the planetary atmosphere
exposed to open field lines through which particles can escape to
interplanetary space.  The value of these three quantities provides
information about the level of exposure that a habitable planet has to
the erosive effects of stellar wind and the potentially harmful
effects of the CR.
%REVISED2

The maximum value of the magnetopause field intensity $B_\rom{mp}$ is
estimated by equating the magnetic pressure $P_\rom{mp}=B_\rom{mp}^2/(2
\mu_0)$ and the dynamical stellar wind pressure $P_\rom{sw}$
(eq. \ref{eq:Psw}),
%REVISED2

\beq{eq:BmpPsw}
B_\rom{mp}=(2 \mu_0)^{1/2} P_\rom{sw}^{1/2}
\eeq
%REVISED2

Although the magnetopause fields arise from very complex processes
(Chapman-Ferraro and other currents at the magnetosphere boundary), in
simplified models $B_\rom{mp}$ is parametrized as a multiple of the
planetary field intensity $B_\rom{p}$ as measured at the substellar
point $r=R_S$ \citep{Mead64,Voigt95},
%REVISED2

\beq{eq:BmpMRs}
B_\rom{mp}=2 f_0 B_\rom{p}(r=R_S)=\left(\frac{f_0 \mu_0}{2\pi}\right) \sqrt{2} {\cal M}
R_S^{-3}
\eeq
%REVISED2

where $f_0$ is a numerical enhancement factor of order 1.  We are
assuming here that the dipolar component of the intrinsic field
dominates at magnetopause distances even in slightly dipolar PMF.
Combining equation \ref{eq:BmpPsw} and \ref{eq:BmpMRs} we obtain an
estimate of the standoff distance:
%REVISED2

\beq{eq:Rs}
R_S = \left(\frac{\mu_0 f_0^2 }{8 \pi^2}\right)^{1/6}
\mathcal{M}^{1/3} P_\rom{sw}^{-1/6}
\eeq
%REVISED2

It is important to stress here that the $R_S$ given by
eq. (\ref{eq:Rs}) assumes that the pressure exerted by the gasses
trapped inside the magentosphere cavity is negligible.  This is a good
approximation only if the planetary magnetic field is very intense or
the stellar wind is weak or the planetary atmosphere is not too
bloated by the XUV radiation.  In the case when any or none of these
condition are fulfilled we will refer to the $R_S$ estimated with
eq. (\ref{eq:Rs}) as the {\it magnetic standoff distance}.
%REVISED2

The last but not least important property in which we are interested
is the area of the polar cap.  The polar cap is the region in the
magnetosphere where open field lines could transport ions into or from
the interplanetary space.  \citet{Siscoe75a} showed that the area of
the polar cap $A_\rom{pc}$ scales with the dipole moment and the
dynamical pressure of the stellar wind as:
%REVISED2

\beq{eq:Apc}
A_\rom{pc}=4.63\times 10^{-2} \left(\frac{\cal M}{{\cal
    M}_\oplus}\right)^{-1/3}
  \left(\frac{P_\rom{sw}}{P_\rom{sw\odot}}\right)^{1/6}
\eeq
%REVISED2

where ${\cal M}_\oplus=7.768\times 10^{22}\,\rm{A m}^2$ and
$P_\rom{sw\odot}=2.24\,\rm{nPa}$ are the present values of the dipole
moment of the Earth and the average dynamic pressure of the solar wind
as measured at the Earth distance \citep{Stacey92,Griebmeier05}.
%REVISED2

In order to model the evolution of the magnetosphere properties we
need to calculate reliable values of the surface dipolar component of
the PMF $B_\rom{p}^{dip}$, the average number density $n$, the
velocity $v_\rom{eff}$ and the temperature $T$ of the stellar wind.
These quantities depend in general on time and also on different
planetary and stellar properties.  In the following sections we
describe the models used in this work to calculate the evolving values
of these critical quantities.
%REVISED2

%============================================================
\subsection{Planetary thermal evolution}
\label{subsec:ThermalEvolution}
%============================================================

We assume here that the main source of a global PMFs in TPs is the
action of a dynamo powered by convection in a liquid iron core.  Other
possible sources of PMFs, dynamo action in a mantle of ice, water or
magma or stellar induced magnetic fields, are not considered here.
%REVISED2

The properties and evolution of a core dynamo will depend on the
internal structure and thermal history of the planet.  Two recent
works have built and solved detailed interior and thermal evolution
models intended to study the generation of intrinsic PMF in super
Earths \citep{Gaidos10,Tachinami11} .  These works have paid attention
to different and complimentary aspects of the problem: while
\cite{Gaidos10} have concentrated on the thermodynamics of the core,
\cite{Tachinami11} have developed a detailed treatment of mantle
rheology and convection.  From these works a robust picture of the
thermal evolution of SEs is starting to arise.  More recently
\cite{Zuluaga12} have shown how by combining the results of these
thermal evolution models and the rotational properties of the planets
it is possible to predict not only the mean intensity of the magnetic
field, but also its regime (dipolar or multipolar).  This last
property is important in predicting the intensity of the dipolar
component of the field at the planetary surface and from there the
dipole moment of the planet.
%REVISED2

\begin{figure}
  \centering
   \includegraphics[width=0.45
     \textwidth]{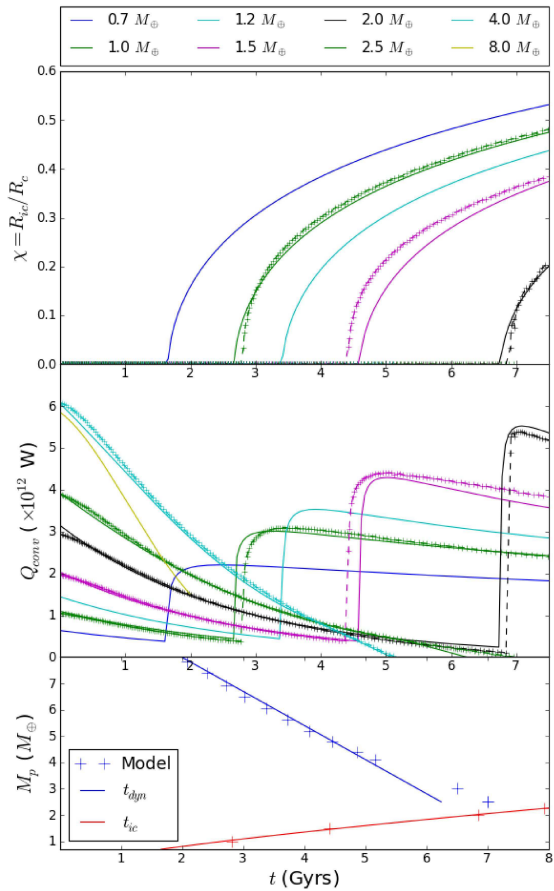}
  \caption{Results of the thermal evolution model by \citet{Gaidos10}
    for TPs with the same composition as the Earth.  Upper panel,
    radius of the solid inner core.  Middle panel, total convective
    power.  Lower panel, time of the inner core nucleation (red) and
    dynamo lifetime (blue).  Numerical results are indicated as
    crosses joined by dashed lines.  Continuous lines correspond to
    the analytical fit used in this work to interpolate and
    extrapolate the model results (see
    text).\vspace{0.2cm}} \label{fig:ThermalEvolution}
    %REVISED2
\end{figure}

There are three key properties that should be predicted by any thermal
evolution model in order to calculate the magnetic properties of the
planet: 1) the total convective power $Q_\rom{conv}$ providing the
energy to be dissipated through dynamo action, 2) the radius of the
solid inner core $R_\rom{ic}$ and thus the height $D$ of the
convecting shell where the dynamo action takes place
($D=R_c-R_\rom{ic}$) and 3) the total dynamo life-time $t_\rom{dyn}$.
Figure \ref{fig:ThermalEvolution} shows the values of these quantities
as predicted by the model by \citet{Gaidos10} for planets with the
same composition as the Earth in the mass range 1.0-4.8 $M_\oplus$.
%REVISED2

%FFFFFFFFFFFFFFFFFFFFFFFFFFFFFFFFFFFFFFFFFFFFFFFFFFFFFFFFFFFFFFFFFFFFF
%FIGURE 1
%FFFFFFFFFFFFFFFFFFFFFFFFFFFFFFFFFFFFFFFFFFFFFFFFFFFFFFFFFFFFFFFFFFFFF
%FFFFFFFFFFFFFFFFFFFFFFFFFFFFFFFFFFFFFFFFFFFFFFFFFFFFFFFFFFFFFFFFFFFFF

We have selected the model results obtained for planets with mobile
lids and habitable surface temperatures $T_\rom{surf}$=288 K.  We
assume that the potentially habitable planets described with our
magnetic protection model fulfill this condition at least for the time
during which they have a dense enough atmosphere and an available
inventory of water.
%REVISED2

In this model all planets start their thermal history with a liquid
iron core with radius $R_c$.  The iron core cools down through the
release of heat stored during the early phases of the formation of the
planet (sensible heat) and generated by the decay of radionuclides.
The contribution of tidal heat in this process has, however, been
negelected.  When the temperature at the center of the planet has
decreased below a critical level, an inner solid core starts to form
and suddenly new sources of energy arise (release of latent heat and
buoyancy generated by the release of light elements).  This change
produces a strong ``rebounce'' of $Q_\rom{conv}$ after the initiation
of inner core growing (see middle panel in figure
\ref{fig:ThermalEvolution}).  However, this phenomenon is restricted
to planets with masses $M_p\lesssim M_\rom{crit}\approx
2-2.5\,M_\oplus$ (a threshold predicted independently by
\citealt{Tachinami11}).  On planets with a larger mass than this limit
the dynamo shuts down before the inner core starts to grow, either
because iron snow is formed at the core mantle boundary (CMB) or the
temperature contrast through the CMB falls below the value required to
ensure convection.
%REVISED2

The model by \citealt{Gaidos10} is restricted to a discrete set of
masses between 1 and 4.8 $\ME$.  In order to interpolate the thermal
evolution properties to other planetary masses we have developed
analytical fits for $R_\rom{ic}$, $Q_\rom{conv}$ and $t_\rom{dyn}$.
We have found that $R_\rom{ic}$ and $Q_\rom{conv}$ could be fitted
using a combination of double exponential functions $f(x)\sim
\exp(x^a)$.  It has been observed that other complex thermodynamic
phenomena could be fitted well by double exponential functions.  This
could imply that our analytical fits more than having a mere practical
value could also have some phenomenological roots.  
%REVISED2

The rising of $R_\rom{ic}$ for planets below the critical mass
$M_\rom{crit}=2\,M_\oplus$ could be computed analytically using:
%REVISED2

\beq{eq:Ric_fit}
\chi(t)=
    {\chi}_{\infty}\left(1-\exp\left[-(t-t_\rom{ic})^{1/4}\right]\right)^3.
\eeq
%REVISED2

Here $\chi_{\infty}$ is a fitting parameter that depends on planetary
mass.  The exponents 1/4 and 3 are found almost independently of
planetary mass.  We have additionally found that $\chi_\infty$ scales
with planetary mass $M_p$ follow a simple power law,
%REVISED2

\beq{eq:ChiScaling}
\chi_\infty(M_p)=1.008\,(M_p/M_\oplus)^{-7/5},
\eeq
%REVISED2

where we have approximated the exponent to the nearest rational value
(-7/5). Using eq. (\ref{eq:Ric_fit}) and (\ref{eq:ChiScaling}) we can
calculate analytical approximations of the inner core radius for any
planetary mass and at any given time after the inner core nucleation
(solid lines in the upper panel of figure \ref{fig:ThermalEvolution}).
As expected the combined fit of $\chi_\infty$ and $\chi(t)$ have some
discrepancies with the numerical results. However, we have verified
that most of them come from the difference between the actual value of
the time of inner-core nucletion $t_\rom{ic}$ and the fitted value of
this quantity (red line in the lower panel of figure
\ref{fig:ThermalEvolution}).  The observed differences, as measured on
the time axis, are no larger than $0.1$ Gyr, i.e. they are below the
resolution of the magnetosphere evolution model presented here.
%REVISED2

$Q_\rom{conv}$ is more complex.  We have fitted the decaying initial
phase (release of sensible heat) and the rebounce phase after inner
core nucleation (release of combined sensible and latent heat)
independently.  The resulting piecewise fitting function reads as:
%REVISED2

\beq{eq:Qconv_fit}
\frac{Q_\rom{conv}}{10^{12}\,{W}}=\left\{
\begin{array}{ll}
q_\rom{Sl}\,\exp(t/t_\rom{Sl})^1,&M<M_\rom{crit}, t<t_\rom{ic}\\
q_\rom{L}\,(t-t_\rom{ic})^{1/3}\times\\
\;\;\times\exp[(t-t_\rom{ic})/t_\rom{L}]^{1/4},&M<M_\rom{crit}, t>t_\rom{ic}\\
q_\rom{Sh}\,\exp[(t-t_\rom{oh})/t_\rom{Sh}]^{2},&M>M_\rom{crit}, \forall\;t
\end{array}
\right.
\eeq
%REVISED2

The fitting parameters $q_\rom{Sl}$, $q_\rom{Sh}$, $q_\rom{L}$,
$t_\rom{Sl}$, $t_\rom{L}$, $t_\rom{oh}$ and $t_\rom{Sh}$ scale also
with planetary mass and the scaling coefficients are presented in
table \ref{tab:ScalingCoefficients}.
%REVISED2

%TTTTTTTTTTTTTTTTTTTTTTTTTTTTTTTTTTTTTTTTTTTTTTTTTTTTTTTTTTTTTTTTTTTTTTTTTTT
%TABLE 1: SCALING COEFFICIENTS 
\begin{table}[ht]
  \centering
  \begin{tabular}{cccc}
    \hline\hline 
    $x$ & $\alpha_{x}$ & $\beta_{x}$ & Mass Range\\
        &          &         & $\ME$ \\\hline
    \hline\multicolumn{4}{c}{Planetary Properties} \\\hline
    $R_p$ & 6378 km & 0.265 & 0.7-10 \\ %CONFIRMED1
    $R_c$  & 1370 km & 0.243 & '' \\ %CONFIRMED1
    $\rho_c$ & 11 g/cm$^3$ & 0.271 & '' \\
    %%%%%%%%%%%%%%%%%%%%%%%%%%%%%%%%%%%%%%%%%%%%%%%%%%%%%%%%%%%%%%%%%%%%%%%
    \hline\multicolumn{4}{c}{$R_\rom{ic}$} \\\hline
    $t_\rom{ic}$ & 2.620 Gyr & 1.360 & '' \\ %CONFIRMED1
    $\chi_{\infty}$ & 1.008 & -0.140 & '' \\ %CONFIRMED1
    %%%%%%%%%%%%%%%%%%%%%%%%%%%%%%%%%%%%%%%%%%%%%%%%%%%%%%%%%%%%%%%%%%%%%%%
    \hline\multicolumn{4}{c}{$Q_\rom{conv}$} \\\hline
    $q_\rom{Sl}$ & 1.090 & 1.528 & 0.7-$M_\rom{crit}$  \\ %CONFIRMED1
    $t_\rom{Sl}$ & 2.907 Gyr & -0.155 & '' \\ %CONFIRMED1
    $q_\rom{L}$ & 13.529 & 1.170 & '' \\ %CONFIRMED1
    $t_\rom{L}$ & 0.190 Gyr & -0.883 & '' \\ %CONFIRMED1
    $q_\rom{Sh}$ & 12.616 & -0.350 & $M_\rom{crit}$-10  \\ %CONFIRMED1
    $t_\rom{oh}$ & -46.096 Gyr & -2.272 & '' \\ %CONFIRMED1
    $t_\rom{Sh}$ & 15.105 Gyr & -0.965 & '' \\ %CONFIRMED1
    \hline\hline
  \end{tabular}
  \caption{Scaling coefficients for planetary properties and thermal
    evolution fitting parameters.  Property $x$ is scaled using a
    simple power law function $x=\alpha_{x}
    (M_p/\ME)^{\beta_{x}}$ where $\alpha_{x}$ is the reference
    value for $x$ (the value at $M_p=1\,\ME$) and $\beta_{x}$ is
    the scaling exponent.  $M_\rom{crit}$ is the maximum mass for
    planets to develop a solid inner core before the shutting down of
    the core dynamo.
  \label{tab:ScalingCoefficients}}
  %REVISED2
\end{table}
%TTTTTTTTTTTTTTTTTTTTTTTTTTTTTTTTTTTTTTTTTTTTTTTTTTTTTTTTTTTTTTTTTTTTTTTTTTT

The combination of analytical fits and power-law scaling of the
fitting parameters allows us also to extrapolate the thermal evolution
results to other planetary masses.  \citealt{Gaidos10} used their
model to study the case of planets with a larger mass than 4.8 $\ME$
focusing on the dynamo lifetime.  On the other hand,
\citealt{Tachinami11} also applied their model for planets with a
smaller mass than the Earth.  We assume here that there are no other
critical phenomena that avoid the extrapolation of the behavior
observed in the reported mass-range to higher and lower planetary
masses.  We will model here planets in the mass range $0.7-10 \ME$.
%REVISED2

Other application of the analytical fitting functions to the thermal
evolution of TPs is that it provides a way to test the impact that
different parameters of the thermal evolution have on the protective
properties of the evolving PMF.  We can check, for example, what would
happen if an improved thermal evolution model predicted a different
critical mass $M_\rom{crit}$ or different values for the time of inner
core nucleation $t_\rom{ic}$.  We will return to this sensitivity
check in section \ref{sec:Discussion}.
%REVISED2

%============================================================
\subsection{Planetary magnetic field}
\label{subsec:ScalingPMF}
%============================================================

In recent years improved numerical experiments have constrained the
full set of possible scaling laws used to predict the properties of
planetary and stellar convection-driven dynamos (see
\citealt{Christensen10} and references therein).  It has been found
that in a wide range of physical conditions the global properties of
convection-driven dynamos can be expressed in terms of simple
power-law functions of the total convective power $Q_\rom{conv}$
available for dynamo action.
%REVISED2

One of the most important results of power-based scaling laws is the
fact that the volume averaged magnetic field intensity inside the
convecting shell $B_\rom{rms}^2=(1/V)\int B^2 dV$ does not depend on
the rotation rate (eq. 6 in \citealt{Zuluaga12}),
%REVISED2

\beq{eq:Brms_scaling}
B_\rom{rms}\approx C_\rom{Brms}\; \mu_0^{1/2} {\bar{\rho}_c}^{1/6} (D/V)^{1/3},
Q_\rom{conv}^{1/3}
\eeq
%REVISED2

where $C_\rom{Brms}=0.24$ for dipolar dominated dynamos and
$C_\rom{Brms}=0.18$ for multipolar dynamos. $\bar{\rho}_c$, $D\approx
R_c (1-\chi)$ and $V\approx 4/3 \pi R_c^3 (1-\chi^3)$ are the average
density, height and volume of the convective shell.  We are assuming
that the whole external liquid iron core is convecting.  In a real
case only a fraction of the core volume is involved in dynamo action
and therefore the magnetic fields predicted with equation
\ref{eq:Brms_scaling} and with our assumption will underestimate the
actual field strenght \citep{Gaidos11}.  We have, however, verified
that this effect is only important for time periods much longer than
the time it takes to start the inner core nucleation.  For a planet
with the same mass as the Earth the time during which the stellar
aggression is the largest is much less than that time for inner core
nucleation (see section \ref{sec:Results}).
%REVISED2

The dipolar field intensity at the planetary surface, and hence the
dipole moment of the PMF, can be estimated if we have information
about the power spectrum of the magnetic field at the core surface.
Although we cannot predict the relative contribution of each mode to
the total core field strength, numerical dynamos exhibit an
interesting property: there is a scalable adimensional quantity, the
local Rossby number $Ro^*_l$, that could be used to distinguish
dipolar dominated from multipolar dynamos.  The scaling relation for
$Ro^*_l$ is (eq. 5 in \citealt{Zuluaga12}):
%REVISED2

\beq{eq:Roml_scaling}
Ro^*_l = C_\rom{Rol}\; {\bar{\rho}_c}^{-1/6} R_c^{-2/3} D^{-1/3} V^{-1/2}
Q_\rom{conv}^{1/2} P^{7/6}.
\eeq
%REVISED2

Here $C_\rom{Rol}=0.67$ is a fitting constant and $P$ is the period of
rotation.  It has been found that dipolar dominated fields arise
systematically when dynamos have $Ro^*_l<0.1$.  Multipolar fields
arise in dynamos with values of the local Rossby number close to and
larger than this critical value.  From eq. (\ref{eq:Roml_scaling}) we
see that in general fast rotating dynamos (low $P$) will have dipolar
dominated core fields while slowly rotating ones (large $P$) will
produce multipolar fields and hence fields with a much lower dipole
moment.
%REVISED2

It is important to stress that the almost independence of
$B_\rom{rms}$ on rotation rate, together with the role that rotation
has in the determination of the core field regime, implies that even
very slowly rotating planets, for example those whose rotation is
locked by the action of the tidal effect of its host star (tidally
locked planets), could have a comparable magnetic energy budget to
rapidly rotating planets with similar size and thermal histories.  In
the former case the magnetic energy will be redistributed among other
multipolar modes rendering the core field more complex in space and
probably also in time.  Together all these facts introduce a
non-trivial dependence of dipole moment on rotation rate very
different than that obtained with the traditional scaling laws used by
previous works (see e.g. \citealt{Griebmeier04} and
\citealt{Khodachenko07}).
%REVISED2

Using the value of $B_\rom{rms}$ and $Ro^*_l$ we can compute the {\it
  maximum dipolar component} of the field at core surface.  For this
purpose we use the maximum dipolarity fraction $f_\rom{dip}$ (the
ratio of the dipolar component to the total field strength at core
surface) that for dipolar dominated dynamos $f_\rom{dip}^{max}\approx
1.0$ and for multipolar ones $f_\rom{dip}^{max}\approx 0.35$ (see
\citealt{Zuluaga12} for details).  To connect this ratio to the
volumetric averaged magnetic field $B_\rom{rms}$ we use the volumetric
dipolarity fraction $b_\rom{dip}$ that it is found, as shown by
numerical experiments, conveniently related with the maxium value of
$f_\rom{dip}$ through eq. (12) in \citealt{Zuluaga12},
%REVISED2

\beq{eq:bdip_fdip}
b_\rom{dip}^{min}=c_\rom{bdip} {f_\rom{dip}^{max}}^{-11/10}
\eeq
%REVISED2

where $c_\rom{bdip}\approx 2.5$ is again a fitting constant.  Finally by
combining eqs. (\ref{eq:Brms_scaling}-\ref{eq:bdip_fdip}) we can
compute an upper bound to the dipolar component of the field at the
CMB: 
%REVISED2

\beq{eq:Bdipmax_scaling}
B_{c}^{dip}\lesssim \frac{1}{b_\rom{dip}^{min}} B_\rom{rms} = 
\frac{{f_\rom{dip}^{max}}^{11/10}}{c_\rom{bdip}}\;B_\rom{rms}
\eeq
%REVISED2

The surface dipolar field strength is estimated using,
%REVISED2

\beq{eq:Bdip_surface}
B_p^{dip}(R_p)=B_c^{dip}\left(\frac{R_p}{R_c}\right)^3
\eeq
%REVISED2

and finally the total dipole moment is calculated using
eq. (\ref{eq:M}) for $r=R_p$.
%REVISED2

It should be emphasized that the surface magnetic field intensity
determined using eq.  (\ref{eq:Bdip_surface}) overestimates the actual
PMF dipole component.  The actual field could be much more complex
spatially.  Using our model we can only predict the maximum level of
protection a given planet could have from a dynamo-generated intrinsic
PMF.
%REVISED2

\begin{figure}  
  \centering
   \includegraphics[width=0.45
     \textwidth]{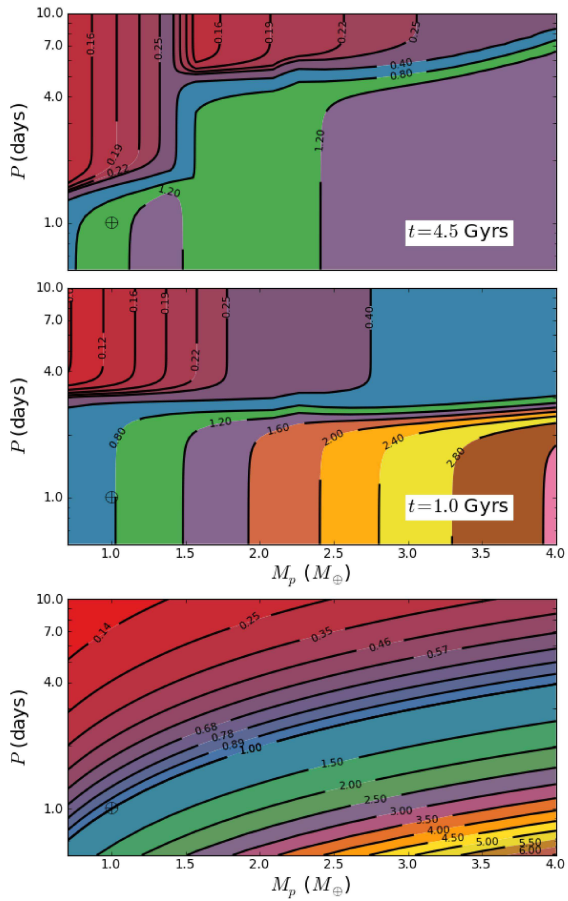}
  \caption{Maximum dipole moment predicted with the power-based
    scaling laws used in this work (upper and middle panels) and the
    rotation dependent scaling law by \citealt{Sano93} (lower
    panel).\vspace{0.2cm}} \label{fig:DipoleMoment}
  %REVISED2
\end{figure}

We show in figure \ref{fig:DipoleMoment} the result of applying the
previously described method to calculate the dipole moment for TPs.
We compare them with the static value of the same quantity as computed
by the rotation dependent dynamo scaling law by \citealt{Sano93} that
has been commonly used in previous works.  The differences between
both approaches are significant.  Not only the estimated values of the
dipole moment are quite different but the general dependence on
planetary mass and rotation period is much more rich and complex.
Those differences have important and previously unknown consequences
in the magnetic protection of tidally locked and unlocked habitable
planets.  We will return to this point in section \ref{sec:Results}.
%REVISED2

%FFFFFFFFFFFFFFFFFFFFFFFFFFFFFFFFFFFFFFFFFFFFFFFFFFFFFFFFFFFFFFFFFFFFF
%FIGURE 2
%FFFFFFFFFFFFFFFFFFFFFFFFFFFFFFFFFFFFFFFFFFFFFFFFFFFFFFFFFFFFFFFFFFFFF
%FFFFFFFFFFFFFFFFFFFFFFFFFFFFFFFFFFFFFFFFFFFFFFFFFFFFFFFFFFFFFFFFFFFFF

%============================================================
\subsection{Stellar properties}
\label{subsec:StellarProperties}
%============================================================

Once we have determined the PMF intensity, the other element required
to estimate the magnetosphere properties is the dynamical pressure of
the stellar wind.  However, magnetic protection does not only depend
on the size of the magnetosphere.  The scale height of the atmosphere
should be also estimated and compared with the magnetopause distance.
Highly irradiated atmospheres, such as those of close-in planets at
early phases of the stellar evolution, could be expanded enough to be
exposed to the direct action of stellar wind.  In this case we will
therfore also need to estimate the level of high energy flux at the
top of the atmopshere of our habitable planets.  In this section we
describe the models used in this work to calculate all the relevant
properties concerning the interaction between the star and the
planetary magnetosphere and atmosphere.
%REVISED2

The properties of low-mass stars are still very uncertain (see
e.g. \citealt{Engle11}).  However, since our magnetic field model is
able only to determine the maximum intensity of the PMF we will only
be interested in limits of the stellar properties providing the lower
level of ``stellar aggression'' (the minimum stellar wind pressure and
XUV irradiation).  Combining upper bounds for the magnetic properties
of the planet and lower bounds for the stellar aggression will give us
an overestimation of the overall magnetic protection of the planet.
In this way if, under our conservative model, a planet results
endangered or unprotected, the actual case will be even worse.
%REVISED2

%......................................................................
\subsubsection{Habitable zones and tidally locking limits}
\label{subsubsec:HZTL}
%......................................................................

The effects a star has on the planetary environment depend on three
basic factors: 1) the fundamental properties of the star, luminosity
$L_\star$, effective temperature $T_\rom{\star}$ and radius $R_\star$,
2) the average distance of the planet from the star (distance to the
HZ) and 3) the relation between the rotation and orbital period, in
particular if they are conmensurable (tidally locked planet) or not5D
(unlocked planet).
%REVISED2

For the stellar properties we are using the results obtained by
\citealt{Baraffe98} (thereafter BCAH98) who calculated the evolution
of low-mass stars in a wide range of metallicities including the solar
value.  In all cases we assume the same metallicity for the stars as
the Sun.  For simplicity we are also assuming that the basic stellar
properties, temperature, luminosity and radius, are constant during
the range of time in which we are calculating the magnetosphere
properties $t=0.7-3$ Gyr (see section \ref{subsubsec:StellarWind}).
To be consistent with the purpose of computing the best case for
magnetic protection, we take the stellar properties predicted by the
model at the end of the time interval, i.e. $t=3.0$ Gyr.  At that time
the HZ will be placed at the greatest distance from the star
especially for stars in the mass range $M_\star=0.8-1.0$ and therefore
the stellar wind pressure and XUV irradiation will be minimum.
%REVISED2

To determine the distance to the HZ we use here the results obtained
by \citealt{Kasting93}.  Although the detailed atmospheric models by
\citealt{Kasting93} were calculated for only three stellar masses, we
use the parabolic fitting developed by \citealt{Selsis07a} to
calculate the inner $a_\rom{in}$ and outer edges $a_\rom{out}$ of the
HZ of arbitrary GKM stars:
%REVISED2

\beq{eq:HZ}
\begin{array}{lll}
a_\rom{in} & = & ( a_{in\odot}-\alpha_\rom{in} \Delta T -
\beta_\rom{in} \Delta T^2 ) \sqrt{L/L_\odot} \\
a_\rom{out} & = & ( a_{out\odot}-\alpha_\rom{out} \Delta T -
\beta_\rom{out} \Delta T^2 ) \sqrt{L/L_\odot}
\end{array}
\eeq
%REVISED2

where $a_{in\odot}$ and $a_{out\odot}$ are the inner and outer edges
of the HZ for the present sun, $\Delta T=T-T_\odot$ and $\alpha$ and
$\beta$ are fitting constants.  To define our HZ we use the
conservative limits given by the criteria of ``recent Venus'' and
``early Mars'' \citep{Kasting93}.  In this case $a_{in\odot}=0.72$,
$\alpha_\rom{in}=2.7619\times 10^{-5}$, $\beta_\rom{in}=3.8095\times
10^{-9}$, $a_{in\odot}=1.77$, $\alpha_\rom{out}=1.3786\times 10^{-4}$,
$\beta_\rom{out}=1.4286\times 10^{-9}$ \citep{Selsis07a}.
%REVISED2

Planets at close-in orbits are affected by the tidal interaction with
the host star.  This interaction dampens the primordial rotation of
the planets leaving them in a resonant rotational state where the
period of rotation $P$ becomes commensurable with the orbital period
$P_o$,
%REVISED2

\beq{eq:P-resonance}
P:P_o=n:2
\eeq
%REVISED2

Where $n$ is an integer larger than or equal to 2.  The value of $n$
is determined by multiple dynamical factors one the most important
being the orbital eccentricity.  The maximum distance $a_\rom{tid}$ at
which a solid planet in a circular orbit becomes tidally locked before
a given time $t$ is given by \citep{Peale77},
%REVISED2

\beq{eq:TidalLimit}
a_\rom{tid}(t) =
0.5\,\rm{AU}\,\left[\frac{(M_\star/M_\odot)^2P_\rom{prim}}{Q}\right]^{1/6}
t^{1/6}
\eeq
%REVISED2

Where the primordial period of rotation $P_\rom{prim}$ should be
expressed in hours, $t$ in Gyr and $Q$ is an adimensional dissipation
function.  For the purposes of this work we assume the same value of
the estimated primordial rotation period of the Earth for all planets,
$P_\rom{prim}\approx 17\,\rm{hours}$ \citep{Varga98,Denis11} and a
dissipation function $Q=100$ typical for terrestrial planets.
%REVISED2

\begin{figure}  
  \centering
   \includegraphics[width=0.45
     \textwidth]{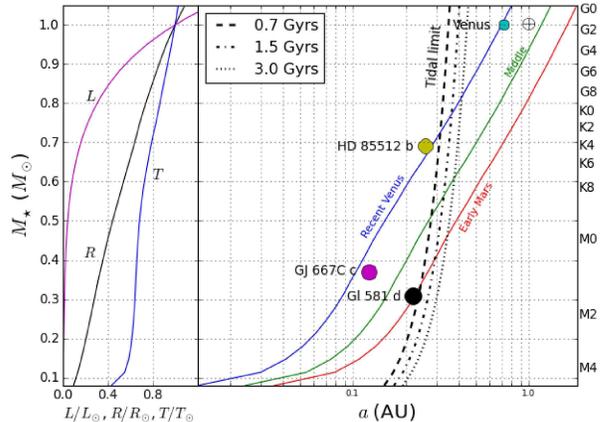}
  \caption{Stellar properties, HZ and tidally locking limits for solar
    metallicity GKM stars. Stellar properties are computed from the
    BCAH98 model at $t=3$ Gyr.  Tidally locking limits assume
    $P_\rom{prim}=17$ hours, and
    $Q=100$.\vspace{0.2cm}} \label{fig:StellarProperties}
  %REVISED2
\end{figure}

In figure \ref{fig:StellarProperties} we depict the value of the
stellar properties, HZ and tidally locked limits for the
stars in the mass-range studied in this work.
%REVISED2

%FFFFFFFFFFFFFFFFFFFFFFFFFFFFFFFFFFFFFFFFFFFFFFFFFFFFFFFFFFFFFFFFFFFFF
%FIGURE 3
%FFFFFFFFFFFFFFFFFFFFFFFFFFFFFFFFFFFFFFFFFFFFFFFFFFFFFFFFFFFFFFFFFFFFF
%FFFFFFFFFFFFFFFFFFFFFFFFFFFFFFFFFFFFFFFFFFFFFFFFFFFFFFFFFFFFFFFFFFFFF

%......................................................................
\subsubsection{Stellar wind}
\label{subsubsec:StellarWind}
%......................................................................

The properties of the stellar wind change in time and vary with the
distance from the host star.  Dur to this dependence on distance and
for the sake of simplicity we have used here the pure hydrodynamical
isothermal model developed by \citealt{Parker58} (hereafter the {\it
  Parker's model}).  It has been shown that this simple model reliably
predicts the stellar wind properties of stars with periods of rotation
of the same order as the present solar value, i.e. $P\sim 30$ days
\citep{Preusse05}.  For rapidly rotating stars, i.e. young stars
and/or active dM stars, the Parker's model underestimates the stellar
wind properties by a factor up to 2 \citep{Preusse05}.  Given the
dependence of the magnetosphere properties on the stellar wind
pressure (equations \ref{eq:BmpPsw}-\ref{eq:Apc}) the
rotation-independent model will give us values for the magnetopause
fields, standoff distances and polar cap areas, between 10-40\% off
the values obtained with a more detailed model (e.g. the extended
model by \citealt{Weber67}).  Magnetopause fields have the largest
uncertainties ($\sim 40\%$) while standoff distances and polar cap
areas, that as we will also shown are the most critical parameters,
are off by just $\sim 10\%$ (standoff radius will be overestimated
while polar cap areas are underestimated).
%REVISED2

According to the Parker's model the stellar wind average particle
velocity $v$ at distance $d$ from the host star is obtained by
solving the {\it Parker's wind} equation:
%REVISED2

\beq{eq:ParkerEquation}
u^2-\log u=4 \log \rho + \frac{4}{\rho} - 3
\eeq

where $u=v/v_c$ and $\rho=d/d_c$ are a normalized velocity and
distance and $v_c=\sqrt{k_B T/m}$ and $d_c=G\Ms m/(4 k_B T)$ are
respectively the local sound velocity and critical distance where the
stellar wind becomes subsonic.  In this model $T$ is the temperature
of the stellar corona and appears here as the key parameter that
determines the properties of the stellar wind.  The number density $n$
is determined from the velocity at each distance using the continuity
equation,
%REVISED2

\beq{eq:Parker_n}
n=\frac{\dot{\Ms}}{4\pi d^2 v m}
\eeq
%REVISED2

The time dependence of the stellar wind properties is much harder to
estimate.  We have used here the formulas derived by
\citealt{Griebmeier04} and \citealt{Lammer04} and that were originally
based on the observational estimates of the stellar mass-loss rate by
\citealt{Wood02} and the theoretical evolution models by
\citealt{Newkirk80}.  For main-sequence stars and times $t>0.7$ Gyr,
the long-term averaged stellar wind velocity and number density as
measured at a distance of 1 AU is estimated by:
%REVISED2

\beq{eq:vt}
v_{1\rm{AU}}(t)=v_0 \left(1+\frac{t}{\tau}\right)^{\alpha_v}
\eeq
%REVISED2

\beq{eq:nt}
n_{1\rm{AU}}(t)=n_0 \left(1+\frac{t}{\tau}\right)^{\alpha_n}
\eeq
%REVISED2

Where $\alpha_v=-0.43$, $\alpha_n=-1.86\pm 0.6$ and
$\tau=25.6\,\rm{Myr}$ \citep{Griebmeier09}. $v_0=3971$ km/s and
$n_0=1.04\times 10^{11}\,\rm{m}^{-3}$ were estimated assuming present
long-term averages of the solar wind as measured at the distance of
the Earth, i.e.  $n(4.6\,\rm{Gyr},1\,\MSUN,1\,\rm{AU})=6.59\times
10^6\,\rm{m}^{-3}$ and
$v(4.6\,\rm{Gyr},1\,\MSUN,1\,\rm{AU})=425\,\rm{km/s}$
\citep{Schwenn90}.  For times $t<0.7$ Gyr stellar wind models are too
uncertain and the estimates provided by eqs. (\ref{eq:vt}) and
(\ref{eq:nt}) become unreliable \citep{Wood02}.
%REVISED2

\citealt{Griebmeier07} devised a clever way to combine the distance
dependent estimation of the stellar wind properties, given for example
by the Parker's model, with the time variation of the reference number
density and velocity given by eqs. (\ref{eq:vt}) and (\ref{eq:nt}).
For the sake of completeness in this work we summarize the procedure
by \citealt{Griebmeier07} but for further details we refer to section
2.4 of that work.
%REVISED2

The stellar wind properties at time $t$ and distance $d$ for a given
stellar mass are calculated by estimating first the coronal
temperature $T(t)$ for which the velocity obtained with the Parker's
equation evaluated at $d=1$ AU coincides with the reference velocity
obtained by eq. (\ref{eq:vt}) evaluated at time t.  Using the coronal
temperature $T(t)$ the Parker's equation provides the velocity at any
distance from the star at that time.  The number density is obtained
from the continuity equation (\ref{eq:Parker_n}) assuming that the
stellar mass-loss rate scales with the stellar radius as $\dot
M_\star(t)=\dot M_\odot(t)(R_\star/R_\odot)^2$, where the solar
reference mass-loss rate $\dot M_\odot$ is computed using:
%REVISED2

\beq{eq:dotMsun}
\dot M_\odot(t)=4\pi (1\,\rm AU)^2\;m\;
n_{1\rm{AU}}(t)
v_{1\rm{AU}}(t)
\eeq
%REVISED2

\begin{figure}
  \centering
  \includegraphics[width=0.45\textwidth]{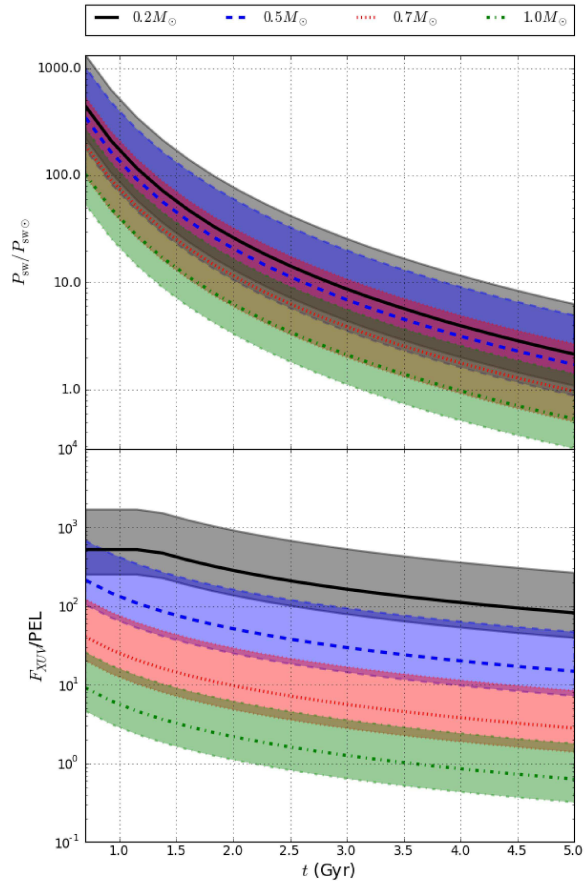}
  \caption{Evolution of the stellar wind dynamical pressure (upper
    panel) and the XUV flux (lower panel) inside the HZ of
    a selected set of stellar masses.\vspace{0.5cm}}
  \label{fig:SW-XUV}
  %REVISED2
\end{figure}

The values of the stellar wind dynamical pressure $P_\rom{dyn}=m n
v^2$ inside the HZ of four different stars as computed using the
procedure described before are plotted in the upper panel of figure
\ref{fig:SW-XUV}.
%REVISED2

%FFFFFFFFFFFFFFFFFFFFFFFFFFFFFFFFFFFFFFFFFFFFFFFFFFFFFFFFFFFFFFFFFFFFF
%FIGURE 4
%FFFFFFFFFFFFFFFFFFFFFFFFFFFFFFFFFFFFFFFFFFFFFFFFFFFFFFFFFFFFFFFFFFFFF
%FFFFFFFFFFFFFFFFFFFFFFFFFFFFFFFFFFFFFFFFFFFFFFFFFFFFFFFFFFFFFFFFFFFFF

%......................................................................
\subsubsection{Stellar XUV fluxes}
\label{subsubsec:XUVFlux}
%......................................................................

The XUV luminosity of a star depends on its level of chromospheric and
coronal activity which in turn depend on the rotation rate of the
star.  It is now well known that the rotation of main sequence stars
slows down with age.  It follows from there that the XUV luminosity
should also decrease monotonically in time.  Estimating the variation
in time of the XUV luminosity for stars of different spectral types is
harder than thought.  There is a large dispersion of rotation periods
and hence in the XUV luminosity of stars of the same spectral type
\citep{Pizzolato03}. Differences in one order of magnitude have been
observed in stars with the same age (see e.g. \citealt{Micela96}).
Additionally EUV radiation is absorbed by the interstellar medium and
therefore we depend on proxies to estimate the total XUV luminosity,
X-ray luminosity being the most common.
%REVISED2

Using observational data and independent estimations of stellar ages
several authors have developed empirical laws providing the value as a
function of time of different XUV luminosity proxies, e.g. $L_X$ (see
e.g. \citealt{Ribas05}, \citealt{Penz08a}, \citealt{Penz08b},
\citealt{Lammer09}, \citealt{Garces11}).  Given the implicit
uncertainties in the estimation of $L_\rom{XUV}$ and to be consistent
with our goal of obtaining the best case for magnetic protection, we
have selected the model predicting the lowest values of the XUV
luminosities.  The empirical law obtained by \citealt{Garces11} is the
best suited for that purpose.  According to that result the X-ray
luminosity of GKM stars change over time following the simple
power-law function:
%REVISED2

\beq{eq:LXFunc}
L_X=\left\{
\begin{array}{ll}
6.3\times 10^{-4} L_\star & \rm{if}\;t<t_i \\
1.89\times 10^{28}\;t^{-1.55} & \rm{otherwise}
\end{array}
\right.
\eeq
%REVISED2

where $L_\star$ is the bolometric luminosity of the star and $t_i$ is
the end of the saturation phase that scales with $L_\star$ as,
%REVISED2

\beq{eq:ti}
t_i=0.06\,\rm{Gyr}\;\left(\frac{L_\star}{L_\odot}\right)^{-0.65}
\eeq
%REVISED2

Luminosities are in units of erg s$^{-1}$. For simplicity we assume
$L_\rom{XUV}\approx L_\rom{X}$.  Using this model the Present Earth
Value of the XUV flux (thereafter {\it PEL}) is 0.64 erg cm$^{-2}$
s$^{-1}$ which, as expected, underestimates the observed value of this
quantity \citep{Judge03,Guinan09}.
%REVISED2

In figure \ref{fig:SW-XUV} we plot the value of the XUV flux as a
function of time as measured in the HZ of four different stars.  XUV
fluxes in the range of $10-1000$ PEL seem to be common in early phases
of stellar evolution posing severe constraints on the survival of
planetary atmospheres of unmagnetized and weakly magnetized planets.
We will return to this point in section \ref{sec:Results}.
%REVISED2

%......................................................................
\section{Atmospheric thermal expansion and mass-loss}
\label{sec:AtmosphericEscape}
%......................................................................

One of the most critical points regarding the survival of the
atmospheres of low-mass planets is the fact that in the early phases
of stellar evolution they would be under extreme conditions of X and
EUV irradiation.  A highly irradiated and relatively light atmosphere
will have high levels of thermal and non-thermal mass-losses
especially if it is unprotected against the action of the stellar
wind.  It has been estimated that under no or even a weak protection
from an intrinsic magnetic field a TP could lose its atmosphere or
most of its volatile content in a time-scale much shorter than that
required for the evolution of life (see e.g. \citep{Lammer12}).
Understanding the relation between the XUV-induced atmospheric
expansion and mass-loss, and the properties of the planetary
magnetosphere is of fundamental importance in assessing the problem of
magnetic protection.
%REVISED2

The key property in distinguishing between a magnetically protected
atmosphere and an exposed one is the radius of the exobase
$R_\rom{exo}$, defined as the distance where the mean-free path of
atmospheric particles could be comparable to the size of the planet.
This is the limit where atmospheric particles, given the proper
energetic or flux conditions, could escape from the planetary
atmosphere.  The height of the exobase depends on many complex factors
ranging from the opacity of the atmopsheric gasses to the high energy
radiation from the star, the interaction between the charged and
neutral components of the high atmosphere and a complex network of
chemical and photochemical reactions of the atmospheric constituents
(for a complete description see \citealt{Tian08}).  All these factors
are critically determined by the chemical composition of the
atmosphere.  In the last few years several authors have, using
detailed chemical, thermal and hydrodynamical models of TPs
atmospheres, calculated the exosphere properties for two chemical
compositions: N$_2$ rich or Earth-like composition atmospheres
\citep{Watson81,Kulikov06,Tian08} and dry CO$_2$ or Venus-like
composition atmospheres \citep{Tian09,Lammer12}.
%REVISED2

\begin{figure}
  \centering
  \includegraphics[width=0.45\textwidth]{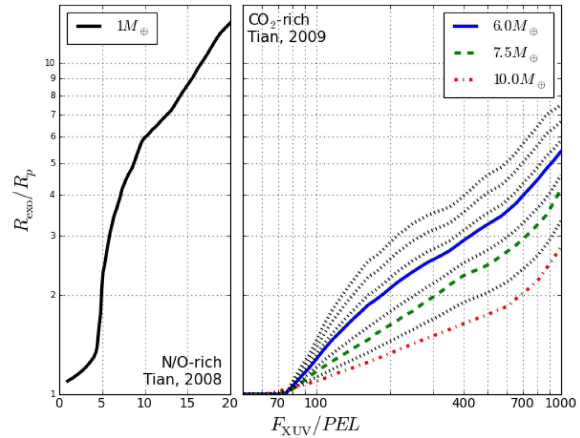}
  \caption{Radius of the exobase for the models used in this work.
    Left panel corresponds to an Earth-mass planet with a N/O-rich
    atmosphere.  Right panel shows the exobase radius for super-Earths
    with CO$_2$-rich atmospheres.  Dotted lines in the right panel are
    the linearly interpolated and extrapolated exobase radius for
    masses from top to bottom 3.5, 4.5, 5.5, 7.0 and 9.0 $\ME$.  XUV
    fluxes are expressed in terms of the present Earth level,
    $PEL\approx 5$ mW/m$^2$.\vspace{0.5cm}}
  %REVISED2
  \label{fig:ExoBaseRadius}
  %REVISED2
\end{figure}

In order to include this important component in our comprehensive
model we have used here the exospheric properties computed with two
different models: a N$_2$ rich atmosphere of an Earth-like planet
\citep{Tian08} and a CO$_2$ rich atmosphere of massive SEs $M_p>6 \ME$
\citep{Tian09}.  The radius of the exobase in these models as a
function of the XUV flux is depicted in figure
\ref{fig:ExoBaseRadius}.  We see that N$_2$ rich atmospheres of
Earth-like planets expand further than $\sim 10 R_p$ at XUV fluxes
larger than $\sim 20$ PEL.  In this case the atmosphere become exposed
even if they have the maximum level of magnetic protection,
i.e. $R_S\sim 10 R_p$ and could be stripped off by thermal and
(stellar-wind related) non-thermal mass-losses.  On the other hand,
CO$_2$ rich atmospheres are cooled off efficiently through the
emission of IR-radiation in the 15 $\mu$m CO$_2$ band and become able
to withstand higher XUV levels.
%REVISED2

%FFFFFFFFFFFFFFFFFFFFFFFFFFFFFFFFFFFFFFFFFFFFFFFFFFFFFFFFFFFFFFFFFFFFF
%FIGURE 5
%FFFFFFFFFFFFFFFFFFFFFFFFFFFFFFFFFFFFFFFFFFFFFFFFFFFFFFFFFFFFFFFFFFFFF
%FFFFFFFFFFFFFFFFFFFFFFFFFFFFFFFFFFFFFFFFFFFFFFFFFFFFFFFFFFFFFFFFFFFFF

In order to evaluate if a TP is magnetically protected we will compare
the evolving value of the exobase radius and the standoff distance
computed with our dynamical magnetosphere model.  It is expected that
at the earliest phases of stellar and planetary evolution the exobase
radius would be larger than the standoff distance irrespective of the
existence of an intense early dynamo.  To expect that the
magnetosphere protects the planetary atmosphere from the very
beginning is simply unrealistic.  Therefore the critical property to
evaluate the level of magnetic protection of a given planet is the
time $\Delta t_\rom{exp}$ during which this exposition state is
maintained.  More precisely the total atmospheric mass lost during
this interval $M_\rom{exp}$ will give us the information we require to
determine if the planetary atmosphere can survive the early
``aggression'' of its host star.  We return to these important
properties of the magnetic protection evolution in the next section.
%REVISED2

%%%%%%%%%%%%%%%%%%%%%%%%%%%%%%%%%%%%%%%%%%%%%%%%%%%%%%%%%%%%%%%%%%%%%%%%%
\section{Results}
\label{sec:Results}
%%%%%%%%%%%%%%%%%%%%%%%%%%%%%%%%%%%%%%%%%%%%%%%%%%%%%%%%%%%%%%%%%%%%%%%%%

Combining all the elements of our comprenhensive model we have
calculated the evolving magnetic protection conditions of potentially
habitable TPs.
%REVISED2

We have caclculated and compared the level of magnetic protection for
planets in the mass range $0.7-10\,\ME$ in three different cases: 1)
instantaneous values of the magnetosphere and exosphere properties,
especially at the earliest phases of the thermal evolution ($t\approx
1$ Gyr), 2) evolution in time of the same properties during a
time-scale comparable to the development of complex life ($t=1-3$ Gyr)
and 3) the cummulative atmospheric mass-loss of major constituents in
the same period of case 2.
%REVISED2

We study these three cases for three type of planets: a) hypothetical
planets in the mass-range $0.7-10\,\ME$, b) an Earth-twin, i.e. a
habitable planet with the same mass and period of rotation than the
Earth, but orbiting different type of stars and c) the Earth-like
planets and SEs already discovered inside the HZ of their respective
host stars, including, for reference purposes, Venus and the Earth
itself.  In table \ref{tab:SEs} we summarize the properties of the
planets in the last group.
%REVISED2

%TTTTTTTTTTTTTTTTTTTTTTTTTTTTTTTTTTTTTTTTTTTTTTTTTTTTTTTTTTTTTTTTTTTTTT
\begin{table*}[ht]
  \centering
  \begin{tabular}{ccccccccccc}
    \hline\hline 
    Planet & $M_p(\ME)$ & $R_p(R_{\oplus})$ & a(AU) & $P_o$ (days) & e &
    S-type & $M_\star(M_{\odot})$ & $R_\star(R_{\odot})$ &
    age(Gyr) & tid.locked\\ \hline\hline
     Earth & 1.0 & 1.0 & 1.0 & 365.25 & 0.016 & G2V & 1.0 & 1.0 & 4.56 & No \\
     Venus & 0.814 & 0.949 & 0.723 & 224.7 & 0.007 & G2V & 1.0 & 1.0 & 4.56 & Probably \\
     GJ 667Cc & 4.545 & 1.5* & 0.123 & 28.155 & $<0.27$ & M1.25V & 0.37 & 0.42 & $>2.0$ & Yes \\
     HD 85512b & 3.496 & 1.4* & 0.26 & 58.43 & 0.11 & K5V & 0.69 & 0.53 & $5.6-8.0$ & Yes \\
     Gl 581d & 6.038 & 1.6* & 0.22 & 66.64 & 0.25 & M3V & 0.31 & 0.29 & $4.3-8.0$ & Yes \\
    \hline\hline
  \end{tabular}
  \caption{Properties of already discovered SEs inside the HZ of their
    host stars. For reference purposes the properties of Venus and the
    Earth are also included.  Values of radii marked with an $*$ are
    unknown and were estimated using the mass-radius relation for
    planets with the same composition as the Earth, i.e. $R_p =
    R_{\oplus} (M_p / \ME)^{0.27}$ \citep{Valencia06}.
    \label{tab:SEs}}
  %REVISED2
\end{table*}
%TTTTTTTTTTTTTTTTTTTTTTTTTTTTTTTTTTTTTTTTTTTTTTTTTTTTTTTTTTTTTTTTTTTTTT

To include the effect of rotation in the properties of the PMF we have
assumed that planets in the HZ of late K and dM stars ($M<0.6
M_\odot$) are tidally locked and therefore their periods of rotation
are equal to their orbital periods (n=2 in eq. \ref{eq:P-resonance}).
For planets that preserve their primordial periods of rotation we
assume values for $P$ in the range $1-100$ days as predicted by models
of planetary formation \citep{Miguel10} with $P=1$ day as the
preferred reference value.
%REVISED2

\begin{figure*}
  \centering
   \includegraphics[width=0.83\textwidth]
                   {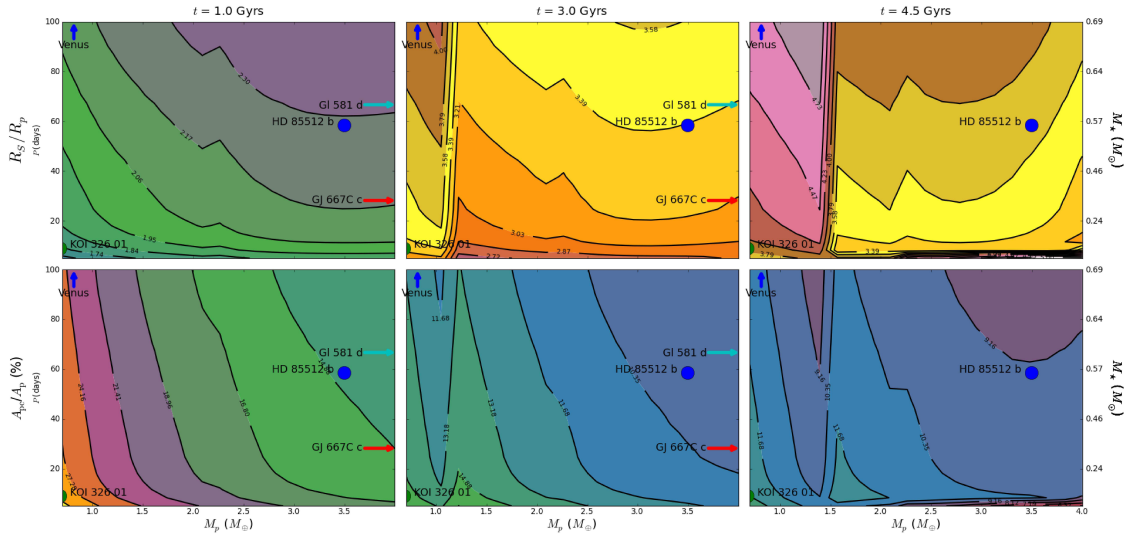}
  \caption{Evolution of the standoff radius and polar cap area for
    tidally locked planets: the period of rotation of each planet is
    assumed equal to the orbital period at the middle of the HZ of low
    mass stars $M_\star<0.6 M_\odot$ (ticks on the right vertical
    axis).  The position of already discovered potentially habitable
    TPs is indicated with a circle or with an arrow when their
    properties fall beyond the axis limits.\vspace{0.5cm}}
  \label{fig:Rs-Apc-Tid}
  %REVISED2
\end{figure*}

\begin{figure*}
  \centering
   \includegraphics[width=0.83\textwidth]
                   {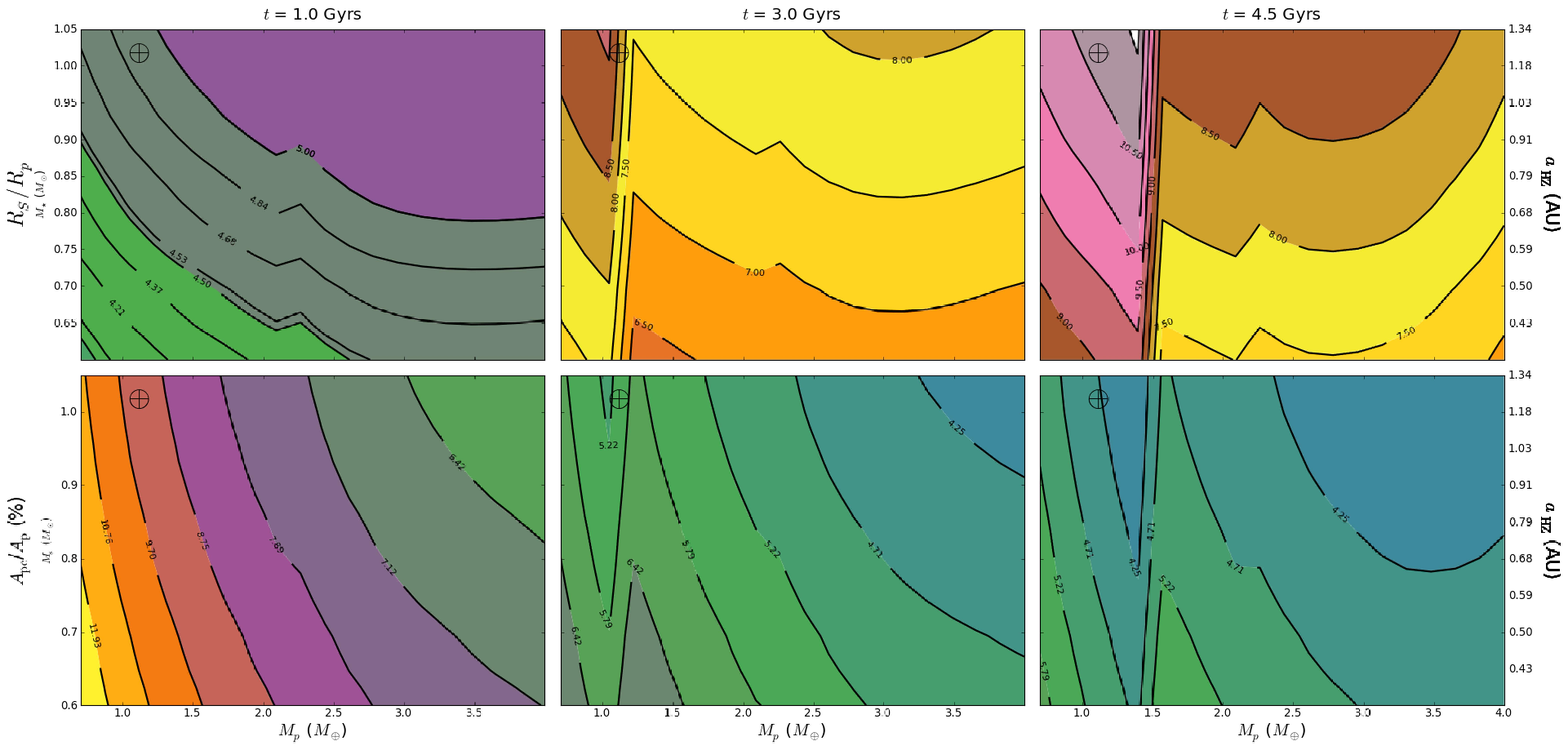}
  \caption{Same as figure \ref{fig:Rs-Apc-Tid} but for unlocked
    planets with constant period of rotation $P=1$ day.\vspace{0.5cm}}
  \label{fig:Rs-Apc-Notid}
  %REVISED2
\end{figure*}

Figures \ref{fig:Rs-Apc-Tid} and \ref{fig:Rs-Apc-Notid} show the
evolution of magnetosphere properties for tidally locked and unlocked
habitable planets.  In all cases we have assumed that the planets are
in the middel of the HZ.  Even in early times tidally locked planets
of arbitrary masses have a non-negligible magnetosphere ($R_S>1.5
\ R_p$).  As the star evolves the dynamical pressure of the stellar
wind decreases more rapidly than the dipole moment (see figures
\ref{fig:DipoleMoment} and \ref{fig:SW-XUV}) and the standoff distance
grows.  The critical boundary observed in the middle and rightmost
panels are a product of the inner core nucleation.  Planets to the
right of the boundary created by a concentration of isolines still
have a liquid core and therefore develop lower dipole moments.  On the
other hand, the inner core in planets to the left of the same boundary
have already started to grow before that time and therefore their
dipole moments are much larger.
%REVISED2

%FFFFFFFFFFFFFFFFFFFFFFFFFFFFFFFFFFFFFFFFFFFFFFFFFFFFFFFFFFFFFFFFFFFFF
%FIGURE 6
%FFFFFFFFFFFFFFFFFFFFFFFFFFFFFFFFFFFFFFFFFFFFFFFFFFFFFFFFFFFFFFFFFFFFF
%FFFFFFFFFFFFFFFFFFFFFFFFFFFFFFFFFFFFFFFFFFFFFFFFFFFFFFFFFFFFFFFFFFFFF

%FFFFFFFFFFFFFFFFFFFFFFFFFFFFFFFFFFFFFFFFFFFFFFFFFFFFFFFFFFFFFFFFFFFFF
%FIGURE 7
%FFFFFFFFFFFFFFFFFFFFFFFFFFFFFFFFFFFFFFFFFFFFFFFFFFFFFFFFFFFFFFFFFFFFF
%FFFFFFFFFFFFFFFFFFFFFFFFFFFFFFFFFFFFFFFFFFFFFFFFFFFFFFFFFFFFFFFFFFFFF

Previous estimates of the standoff radii for tidally locked planets
are lower than the values reported here.  For example,
\citealt{Khodachenko07} place the range of standoff distances well
below $2 \ R_p$ even under mild stellar wind conditions (see figure 4
in their work).  This is easily explained since they also
underestimate the maximum dipole moment for this type of planet.
While they predict maximum dipole moments for tidally locked planets
around stars with $M_\star\approx 0.5 M_\odot$ in the range of 0.022
to 0.15 ${\cal M}_\oplus$, our model predicts dipole moments as large
as 0.8 ${\cal M}_\oplus$ for planets with the largest mass and at
times as early as $t\approx 1$ Gyr.
%REVISED2

A lower standoff distance means a larger polar cap.  Tidally locked
planets, even under the assumption that their exobases are not larger
than the magnetosphere, have more than 15\% of their atmospheric
surface area exposed to open field lines where thermal and non-thermal
processes could efficiently remove atmospheric gasses.  Moreover, our
model predicts that this type of planet will probably have multipolar
PMF (thereafter paleomagnetospheres) which only increase the areas
where field lines are open to the interplanetary space and magnetotail
regions \citep{Stadelmann10}.
%REVISED2

Unlocked planets (figure \ref{fig:Rs-Apc-Notid}) as expected seem to
be best protected by extended magnetospheres $R_S>4 \ R_p$ and lower
polar cap areas $A_\rom{pc}<10\%$.  It is interesting to notice that
in both cases and at early times $t\approx 1$ Gyr a smaller planetary
mass also implies a lower level of protection.  This fact seems to
contradict the idea that low-mass planets ($M_p<2.5$) are best suited
to develop intense PMF \citep{Gaidos10,Tachinami11,Zuluaga12}.  This
apparent contradiction is explained by taking into account that the
early dynamo generated magnetic fields of planets with very different
masses are of the same order (see e.g. figure 8 in
\citealt{Gaidos10}): although larger planets produce more convective
energy, the oversized liquid core and a larger planetary radius
produce a surface magnetic field similar in intensity to that of a
smaller planet.  Having similar surface PMF, planets with larger
masses will have much larger dipole moments, ${\cal M}\sim R_p^3$ and
under the same stellar conditions will be best protected against the
action of the stellar wind.  This situation is not longer valid when
low-mass SEs ($M_p<2.5$) develop an inner core and the available
energy for convection is largely increased.
%REVISED2

We have also calculated the standoff distance and polar cap area for
different evolutionary stages of the already discovered habitable
planets enumerated in table \ref{tab:SEs}.  The position are indicated
with circles, whose size is proportional to their measured or
estimated planetary radii, in the contour plots of figures
\ref{fig:Rs-Apc-Tid} and \ref{fig:Rs-Apc-Notid}.  Planets whose
properties are out of the ranges used in these figures are indicated
with arrows at the border of each panel.
%REVISED2

\begin{figure}
  \centering
   \includegraphics[width=0.45\textwidth]{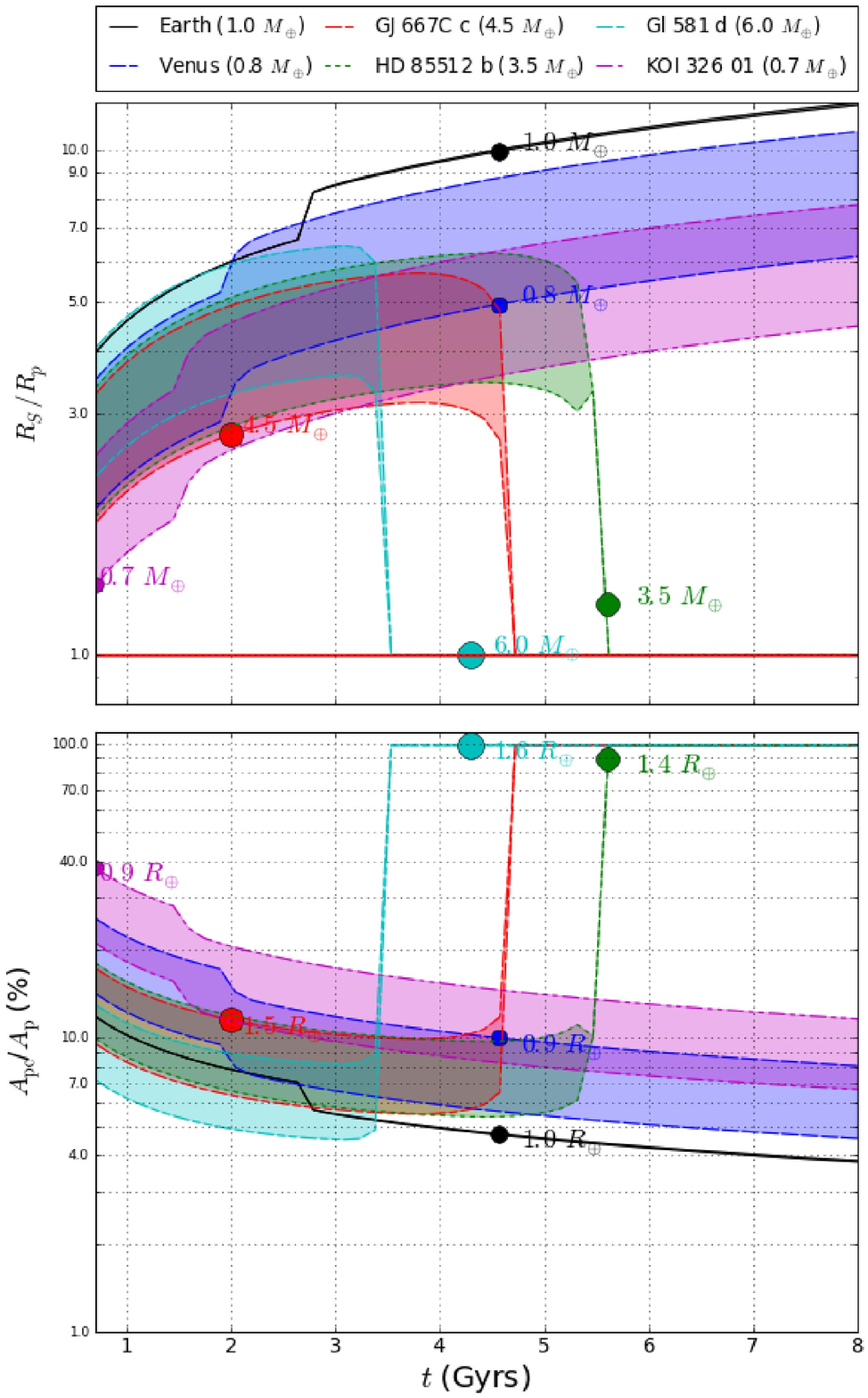}
  \caption{Evolution of magnetosphere properties for already
    discovered habitable SEs.  Shaded regions are limited by the
    properties determined with the minimum ($P\approx 1$ day) and
    maximum period of rotation of the planet ($P\approx
    P_o$).\vspace{0.5cm}}
  \label{fig:Rs-Apc-SEs}
  %REVISED2
\end{figure}

A more detailed account of the evolution of magnetosphere properties
for already discovered habitable planets is presented in figure
\ref{fig:Rs-Apc-SEs}.  The values of magnetosphere properties at the
present age of the planet are indicated with circles whose size is
proportional to the planetary radii. The case of Venus and that of the
Earth are also included for reference purposes.
%REVISED2

%FFFFFFFFFFFFFFFFFFFFFFFFFFFFFFFFFFFFFFFFFFFFFFFFFFFFFFFFFFFFFFFFFFFFF
%FIGURE 8
%FFFFFFFFFFFFFFFFFFFFFFFFFFFFFFFFFFFFFFFFFFFFFFFFFFFFFFFFFFFFFFFFFFFFF
%FFFFFFFFFFFFFFFFFFFFFFFFFFFFFFFFFFFFFFFFFFFFFFFFFFFFFFFFFFFFFFFFFFFFF

In all cases we have used values for the period of rotation ranging
from a minimum, corresponding to an assumed primordial rotation rate
of $P_\rom{min}=24$ h (with exception of the Earth for which we have
used $P_\rom{min}=17$ h), and a maximum, equal to the orbital period
(tidally locking).  The minimum and maximum values of the rotation
period determine the upper (lower) and lower (upper) boundaries of the
shaded regions in the upper panel (lower panel) respectively.
%REVISED2

The case of Venus is particularly interesting in order to analyse the
rest of the planets.  The dynamo of Venus probably shut down at $t=3$
Gyr as a consequence of the drying of the mantle
\citep{Christensen09b}.  A massive loss of water induced by a runaway
greenhouse and unsufficient early magnetic protection played a central
role in the extinction of the early Venusian PMF.  The case of Gl 581
d, GJ 667Cc and HD 85512b, though similar to Venus, is much more
complex.  On one hand their masses are larger than Venus' and
therefore their gravity could provide additional protection to the
atmospheric mass-loss.  On the other hand, their dynamos shut down in
times $t\approx3-5$ Gyr exposing them to the direct action of the
stellar wind.  In G and K stars this situation could not be consider a
big threat since the stellar wind had also decreased its intensity
when the planet lost its PMF (see figure \ref{fig:SW-XUV}), this will
also be the case for HD 85512b.  However, Gl 581d and GJ 667Cc are
located at the HZ of dM stars where the stellar wind pressures, even
at times as late as 4 Gyr, are intense enough to massively erode their
atmospheres or to make them lost their volatile content.
%REVISED2

As explained in section \ref{sec:AtmosphericEscape} the magnetic
protection of a habitable planet is not only a function of the
magnetopshere properties.  In order to asses the problem we need also
to evaluate the effect of the XUV radiation in the outer atmosphere
expansion.  In figure \ref{fig:ExoBaseRadius} we compare the evolution
of the exobase radius and the standoff distance for two representative
TPs planets: an Earth-mass planet with a N/O-rich atmosphere in the HZ
of G-K stars and a super-Earth with $M_p=6 \ME$ and a CO$_2$-rich
atmosphere in the HZ of dM stars $M_\star<0.6$.  For the Earth-mass
planet we have assumed periods of rotation in the range $P=1-10$ days
which are compatible with TP formation theories \citep{Miguel10}.  For
the rotation rate of the super-Earth we have assumed rotation periods
of between 1 day (primordial rotation) and the orbital period (tidally
locked case).  Both planets were placed at the middle of the HZ of
each star.  Standoff distances for the HZ of stars in the mass-range
studied in each case do not change significatively.  For reference
purposes we have only plotted this quantity for the star with the
largest mass in the interval considered in each case (1.0 $M_\odot$
for the upper panel and 0.5 $M_\odot$ for the lower panel)
%REVISED2

%FFFFFFFFFFFFFFFFFFFFFFFFFFFFFFFFFFFFFFFFFFFFFFFFFFFFFFFFFFFFFFFFFFFFF
%FIGURE 9
%FFFFFFFFFFFFFFFFFFFFFFFFFFFFFFFFFFFFFFFFFFFFFFFFFFFFFFFFFFFFFFFFFFFFF
\begin{figure}
  \centering \includegraphics[width=0.45\textwidth]
             {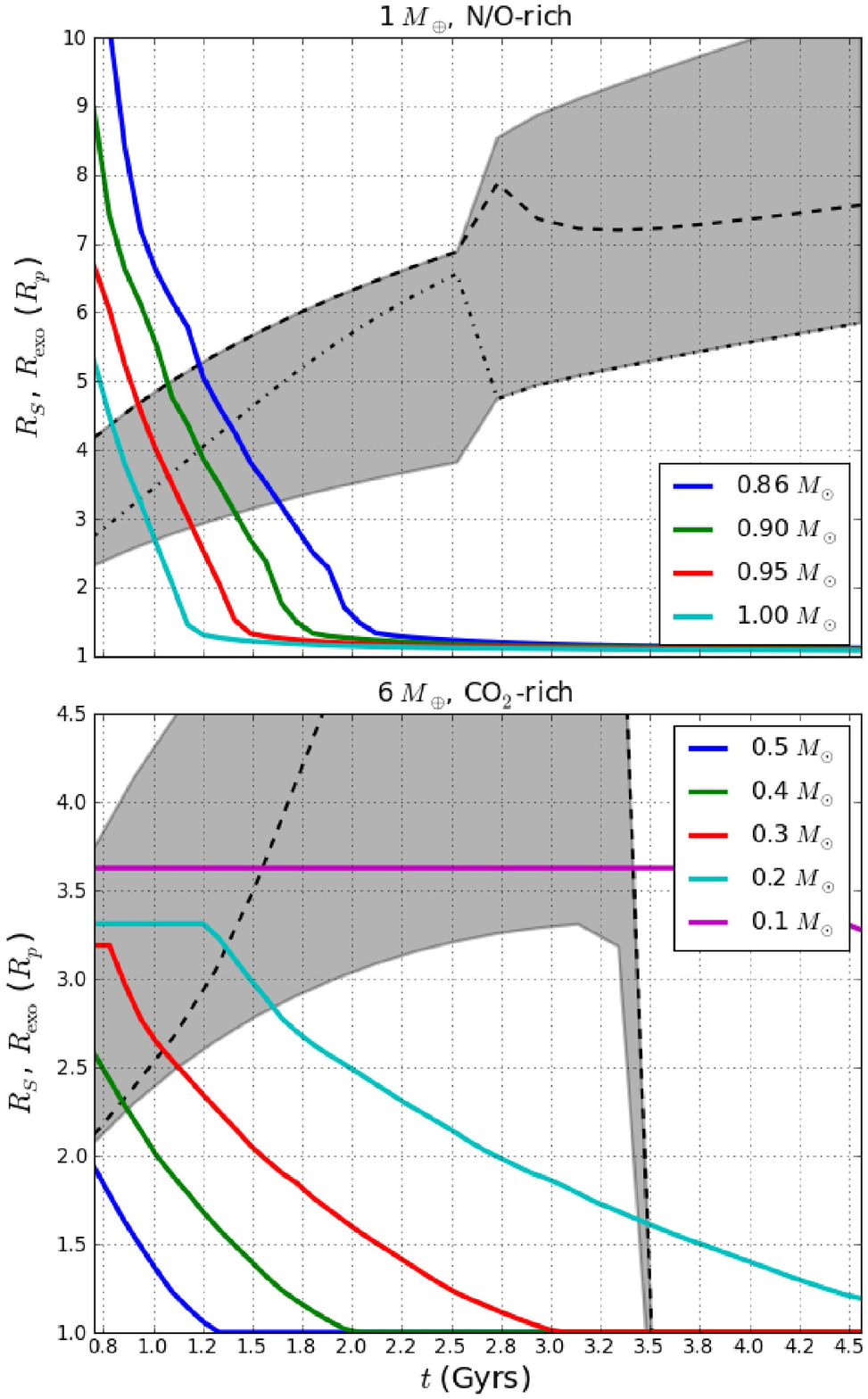}
  \caption{Radius of the exobase (solid lines) and standoff distance
    (shaded area) as a function of time for two potentially habitable
    TPs.  Upper panel: Earth-mass planet with a N/O-rich atmosphere.
    The shaded region corresponds to the standoff distance computed
    for a solar-mass star and assuming periods of rotation of 1 day
    (upper limit) and 10 days (lower limit).  Dashed line is for
    $P=1.5$ days and dash-dotted line is for $P=3$ days. Lower panel:
    massive super-Earth with a CO$_2$-rich atmosphere.  The shaded
    region corresponds to the standoff distances calculated in the HZ
    of a star with $0.5 M_\odot$ and periods between 1 day (upper
    limit) and the orbital period (tidally locked case, lower limit).
    Dashed line corresponds to a rotation period $P=4$ days or
    approximately 1/17th of the orbital period.\vspace{0.5cm}}
  %REVISED2
  \label{fig:ProtectionEvolution}
\end{figure}
%FFFFFFFFFFFFFFFFFFFFFFFFFFFFFFFFFFFFFFFFFFFFFFFFFFFFFFFFFFFFFFFFFFFFF

For a solar-mass star and $P=1$ day (dipolar dominant early PMF), the
exposure time is $\Delta t_\rom{exp}\approx 200$ Myr.  However, if the
primordial period of rotation is 3-5 days (multipolar field) the
exposure could be increased by up to 1 Gyr threatening the atmospheric
stability or its content of volatiles.  The exposure time for an
Earth-mass planet increases monotonically when it is located at closer
distances to late G and early K stars.  If the Earth-mass planet is
located in the middle of the HZ of a star with $M_\star<0.86$
($a_\rom{HZ}\approx 0.80$ AU) it would be subject to levels of XUV
irradiation high enough to blowing off the atmosphere.  This fact
points to the existence of a subregion of the HZ that we can call a
{\it Magnetic-restricted Habitable Zone} (MHZ), where planets with a
given mass and atmospheric composition could preserve their
atmospheres.  It is interesting to note that although Venus was inside
the solar HZ during the first Gyr of the solar system evolution it has
always been outside of the MHZ for Earth-like planets with N/O-rich
atmospheres.  CO$_2$ rich atmospheres which are less propense to
expand, i.e. $R_\rom{exo}<2$ for $F_\rom{XUV}\sim 100$
\citep{Kulikov06} are probably protected at the critical early phases
of planetary evolution, provided the planet preserves the conditions
required for dynamo action (e.g. an hydrated mantle, mobile lids,
etc.)  In that case the MHZ will extend to stars with lower masses or
closer habitable distances.
%REVISED2

The case for massive super-Earths in the HZ of dM stars has several
interesting differences with respect to the magnetic protection of
Earth-mass planets.  The continued activity of very late dM stars
($M_\star\sim 0.1$) will produce a constant level of XUV irradiation
and hence a constant exobase radius during times longer that the
dynamo life-time itself (magenta solid line).  Under this condition
and assuming that the planet is tidally locked in the first Gyr, the
atmosphere will be permanently exposed and will probably be subject to
large thermal and non-thermal mass-losses.  However, for stars with
masses $M_\star\gtrsim 0.2$ a tidally locked planet will only be
exposed during few Myr to 1.0 Gyr.  If the mass-loss is not too large
\citep{Tian09} the planetary atmosphere could survive the initial
stellar aggression.  Again we could identify here a minimum stellar
mass (HZ distance) for which the atmosphere is critically exposed.
For a tidally locked 6 $\ME$ super-Earth that minimum mass is close to
0.3 $M_\odot$ ($a_\rom{HZ}\approx 0.15$ AU).  The inner limit of the
so-called MHZ is farther away for SEs with larger masses but still
covers a significant fraction of the tidally locked region of the HZ.
%REVISED2

\begin{figure}
  \centering \includegraphics[width=0.45\textwidth]
             {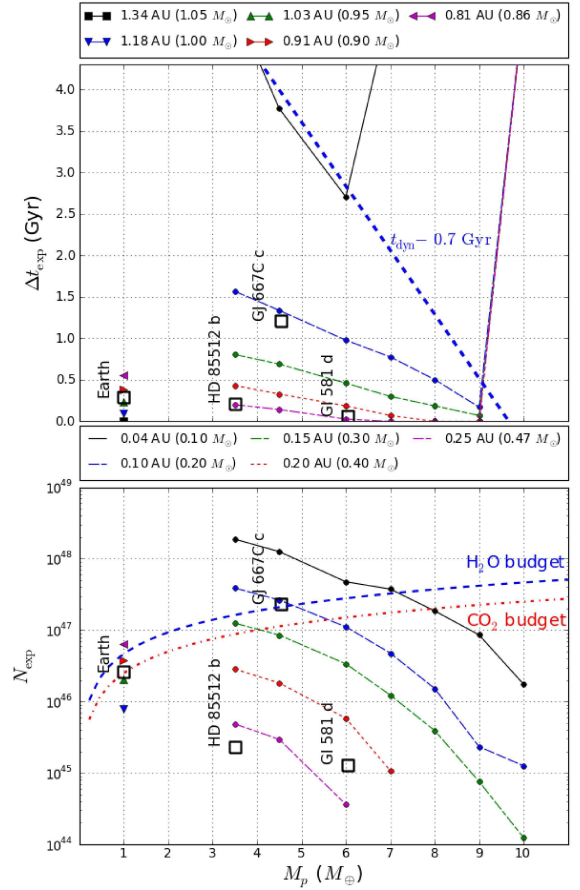}
  \caption{Exposure time $\Delta t_\rom{exp}$ and mass-loss
    $M_\rom{exp}$ as a function of planetary mass for two atmospheric
    compositions: N/O-rich atmospheres (Earth-mass planet) and
    CO$_2$-rich atmospheres (super-Earths).  Planets were placed at
    different distances from their host stars (see legends).  For each
    distance the mass of the star, where that distance corresponds to
    the middle of the HZ, is indicated in parenthesis.  For the
    Earth-mass planet we assumed a constant period of rotation $P=1$
    day.  The super-Earths were placed in the HZ of dM stars and are
    tidally locked, i.e. $P=P_o$.  For reference we have included the
    duration of the dynamo $t_\rom{dyn}$ with respect to the initial
    time (0.7 Gyr) (upper panel, dashed line) and the CO$_2$ and
    H$_2$O content scaled with the Earth's mass (dahed and dash-dotted
    lines in the lower panel).  The exposure time and mass-loss
    predicted for the already discovered habitable planets are marked
    with squares.\vspace{0.5cm}}
  \label{fig:ExposureConditions}
\end{figure}

In order to quantify in more detail the effect that the early
exposition to the stellar wind has on the atmosphere erosion of
habitable TPs we have calculated the exposure time and the total mass
loss during this critical period for an Earth-mass planet in the HZ of
G-K stars and tidally locked SEs. The results are presented in figure
\ref{fig:ExposureConditions}.
%REVISED2

%FFFFFFFFFFFFFFFFFFFFFFFFFFFFFFFFFFFFFFFFFFFFFFFFFFFFFFFFFFFFFFFFFFFFF
%FIGURE 10
%FFFFFFFFFFFFFFFFFFFFFFFFFFFFFFFFFFFFFFFFFFFFFFFFFFFFFFFFFFFFFFFFFFFFF
%FFFFFFFFFFFFFFFFFFFFFFFFFFFFFFFFFFFFFFFFFFFFFFFFFFFFFFFFFFFFFFFFFFFFF

In the case of super-Earths we have used the flux of carbon reported
in figure 8 of \citealt{Tian09}.  For masses not included in the
simulations we have linearly interpolated and extrapolated the already
reported results for planets with masses of 6, 7.5 and 10 $\ME$.  The
flux of oxygen, the most abundant ion in higly irradiatend N/O-rich
atmospheres, in the case of the Earth-mass planet, was calculated
multiplying the estimated number density $n$ as estimated from the
exobase definition (eq. 18 in \citealt{Kulikov06}) and the exobase
bulk velocity reported in figure 8d of \citealt{Tian08}.  In both
cases the total mass-loss is compared with the assumed initial content
of CO$_2$ ($2.5\times 10^{46}$ for an Earth-mass planet,
\citealt{Tian09}) and the amount of water in the planetary hydrosphere
($4.6\times 10^{46}$ molecules for an Earth-mass planet,
\citealt{Kulikov06}).  For planets with a mass larger than that of the
Earth and for the sake of simplicity these initial content of critical
volatiles were scaled linearly with mass.
%REVISED2

We have observed that exposure time and mass-loss, depend on the
distance from the star but are almost independent of the stellar mass
compatible with that distance.  Therefore, an Earth-mass planet placed
at $\sim 0.8$ AU will be subject to a comparable exposure around a
0.75 $M_\odot$ where that distance correspond to the outer boundary of
the HZ (see figure \ref{fig:StellarProperties}) or around a 0.86
$M_\odot$ star (middle of the HZ) or around a 1.05 $M_\odot$ star
(inner boundary of the HZ).  This is the reason why we have decided to
parametrize the results in terms of the planetary distance from the
star rather than in terms of stellar mass as in figure
\ref{fig:ProtectionEvolution}.
%REVISED2

The exposure times for the Earth-mass planet are compatible with the
results presented in figure \ref{fig:ProtectionEvolution}.  The early
mass-loss for this type of planet ranges from 25\% of the scaled mass
of the ocean for a distance 20\% larger than present Earth, to a value
larger than this critical threshold for a planet 20\% closer than the
Earth.  The exposure time and mass-loss for tidally locked
super-Earths decrease with mass as expected from the decreasing of the
exobase radius with an increasing planetary mass (see figure
\ref{fig:ExoBaseRadius}.  We confirm here the existence of a minimum
distance for a given planetary mass beyond which the carbon mass-loss
is below the expected CO$_2$ content.  For example, a super-Earth with
mass $6 \ME$ located at distances larger than 0.1 AU (long dashed
lines in the lower panel) will always have mass-losses below that
threshold.  This limit corresponds to the middle of the HZ around a
$0.2 M_\odot$ star which confirms the analysis derived from figure
\ref{fig:ExoBaseRadius}.
%REVISED2

With the information at hand and assuming that the already potentially
habitable super-Earths have similar compositions to the Earth and
CO$_2$ rich atmospheres we can conclude that GJ 667Cc has already lost
a significant fraction of its atmospheric mass and it is probably now
uninhabitable.  Gl 581d and HD 85512b seem to be in a safe region of
the parameter space.  Altough our model provides a very conservative
estimate of the early exposure conditions, the estimated mass-losses
for both planets are more than one order of magnitude below the
critical threshold.  Even if we accept that their dynamos are weaker
or if we include the additional mass-loss produced after their dynamos
shut down (see figure \ref{fig:Rs-Apc-SEs}) the total mass-losses are
still below the scaled CO$_2$ content.
%REVISED2

%%%%%%%%%%%%%%%%%%%%%%%%%%%%%%%%%%%%%%%%%%%%%%%%%%%%%%%%%%%%%%%%%%%%%%%%%
\section{Discussion and further analysis}
\label{sec:Discussion}
%%%%%%%%%%%%%%%%%%%%%%%%%%%%%%%%%%%%%%%%%%%%%%%%%%%%%%%%%%%%%%%%%%%%%%%%%

This is the first attempt to integrate into a single comprehensive
model all the relevant physical phenomena involved in the magnetic
protection of habitable planets.  Although several components of the
model are expecting important improvements in the following years
(bulk properties of low-mass stars and their evolution, structure and
evolution of stellar winds, physics of highly irradiated atmospheres,
atmospheric escape in magnetized planets) the esential elements have
been put together to produce an integrated view of the role of
magnetic fields in the survival of the atmosphere of habitable TPs.
%REVISED2

One important source of uncertainties in our model, especially when
applied to already discovered TPs, is the assumption that all of them
have similar compositions to the Earth.  Planets with elemental and
mineralogical compositions different to our planet are probably more
abundant than previously thought \citep{Bond10}.  Although numerical
models of the bulk interior of solid planets with very different
compositions have already been computed \citep{Seager07} the detailed
structure, mineralogical phases, thermal profiles and evolution, among
other geophysical relevant information, are waiting to be studied in
more detail for this type of planet.  Plate tectonics, mantle
rheology, additional interior heating sources (e.g. radioactive and
tidaly heating) and the formation, composition and thermal structure
of a metallic core are key properties to study the thermal and
magnetic field evolution on planets with very different composition to
the Earth.
%REVISED2

\citet{Gaidos10} have studied the effect that an increased core size
(larger Fe/Si ratio) or a different amount of radionuclides have on
the thermal evolution and hence the PMF evolution of planets with
different composition than the Earth.  They found that increasing the
Fe/Si ratio for an Earth-mass planet, i.e. increasing the radius of
the core, will have two effects on the thermal and magnetic field
evolution (see figure 7 in their paper): 1) an earlier formation of
the solid iron core and 2) more intense surface fields (mainly due to
a reduction in the distance between the planetary and core surface
without a significant change in the convective power).  A change in
the predicted magnetic field intensity of almost one order of
magnitude was observed changing the core radius between 0.5 and 1.5
times the Earth's core radius.  Using equations \ref{eq:M} and
\ref{eq:Rs} it is predicted that a larger relative amount of Fe could
imply a standoff distance of up to 3 times larger than that predicted
for a planet with a composition similar to Earth.  Also an earlier
inner core formation would imply increased magnetic protection during
the critical earlier phases of stellar and planetary evolution.  The
effect of a different amount of radionuclides on the PMF strength is
negligible (see figure 7 in \citealt{Gaidos10}).  In all these cases
the magnetic protection conditions predicted here will be improved.
Thus, for example, GJ 667Cc will have better chances of preserving its
atmosphere if its content of Fe is much larger than that of the Earth.
%REVISED2

The effect of different rheological properties and thermal structure
on SEs have also been studied in the detailed mantle and core thermal
evolution by \citet{Tachinami11}.  Although they also assumed the same
elemental and mineralogical composition as the Earth the attention
they put on the role of variations in the rheological properties of
the mantle and different thermal conditions at the core mantle
boundary (CMB) could be helpful in roughly guessing what could happen
on planets with different compositions.  They have found that the
total lifetime and surface strength of the PMF are sensitive to
changes in these properties (see figure 11 in \citealt{Tachinami11}).
Changes become particularly important for planets with masses larger
than the Earth ($M_p>2 \ME$).  For example, they predict that the
metallic core of planets with a mass as large as 5 $\ME$ could
actually cool enough to develop an inner solid core, provided that on
one hand the viscosity of the mantle is weaker dependent on pressure
and on the other hand, that the temperature contrast through the CMB
is up to 10 times larger than that expected for the Earth.  In this
case larger-mass-planets could develop strong PMF and its magnetic
protection could be ensured even at closer distances to their host
star than those predicted here.
%REVISED2

\begin{figure}
  \centering \includegraphics[width=0.45\textwidth]
             {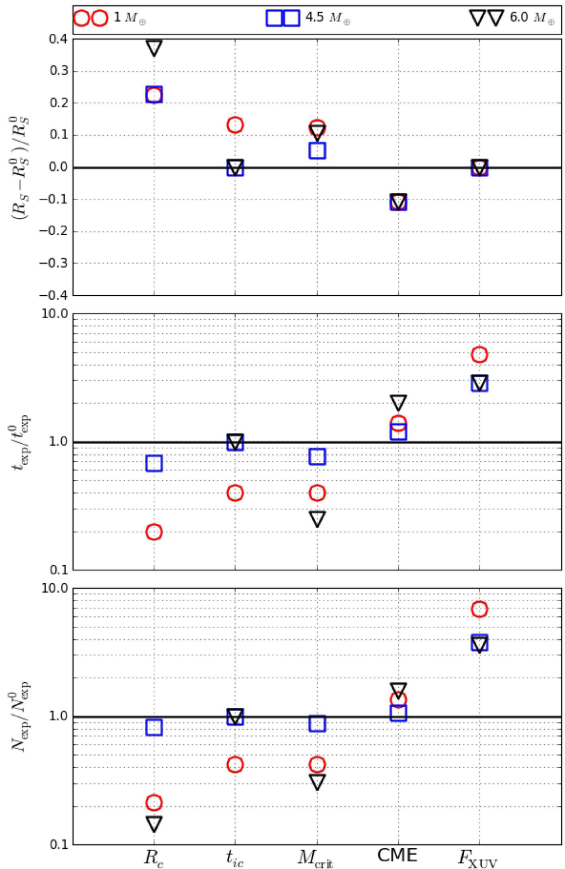}
  \caption{Test of the model's sensitivity to changes in the
    parameters of the thermal evolution model ($R_c$, $t_\rom{ic}$ and
    $M_\rom{crit}$) and the stellar wind and XUV irradiation
    properties (coronal mass ejection conditions or CME and
    $F_\rom{XUV}$).  The value of the magnetic protection key
    quantities $R_S$, $t_\rom{exp}$ and $N_\rom{exp}$ are compared
    with their nominal values $R_S^0$, $t_\rom{exp}^0$ and
    $N_\rom{exp}^0$ as calculated with the set of physical parameters
    used in this work.\vspace{0.1cm}}
  \label{fig:SensitivityTest}
  %REVISED2
\end{figure}

We have simulated the previously discussed effects by changing the
fitting parameters of the phenomenological functions describing the
thermal evolution (equations \ref{eq:Ric_fit}-\ref{eq:Qconv_fit} and
scaling parameters in table \ref{tab:ScalingCoefficients}).  We
present in figure \ref{fig:SensitivityTest} the results of this ``test
of sensitivity'' to our model of variations in the thermal evolution
details.  We have plotted the percentual variation of the standoff
distance and the factor of variation in the total mass-loss when the
following properties of the thermal evolution are modified with
respect the nominal values used in this work: 1) the radius of the
metallic core $R_c$ (scaling parameter $\alpha_\rom{Rc}$), 2) The time
of inner core nucleation $t_ic$ (scaling parameter $\alpha_\rom{tic}$)
and 3) the critical mass for solid inner core nucleation,
$M_\rom{crit}$.
%REVISED2

%FFFFFFFFFFFFFFFFFFFFFFFFFFFFFFFFFFFFFFFFFFFFFFFFFFFFFFFFFFFFFFFFFFFFF
%FIGURE 11
%FFFFFFFFFFFFFFFFFFFFFFFFFFFFFFFFFFFFFFFFFFFFFFFFFFFFFFFFFFFFFFFFFFFFF
%FFFFFFFFFFFFFFFFFFFFFFFFFFFFFFFFFFFFFFFFFFFFFFFFFFFFFFFFFFFFFFFFFFFFF

We see that the value of standoff distance predicted by the model is
very robust. Most of the modification of the model parameters produce
changes in the nominal value of only $\sim$ 20\% or less.  This could
be explained by the weak dependence of the standoff distance on the
dipolar moment, eq. (\ref{eq:Rs}) $R_S\sim{\cal M}^{1/3}$. Only when
the core radius is increased (we have used $R_C\approx 1.2 R_C^0$
where $R_C^0$ is the nominal core radius scaled with the mass of the
Earth) is a difference in the standoff radius of 40\% observed in the
case of massive SEs.  For planets with lower masses the difference is
still around the 20\% limit.
%REVISED2

The exposure time and mass-loss are much more sensitive to changes in
the thermal evolution parameters.  This is explained taking into
account that even small variations in the standoff distance could
increase by a factor of 2-5 the time during which the planet is
exposed directly to the stellar wind erosion.  As expected the largest
changes are observed when we change the core radius (Fe/Si ratio).
The predicted mass-losses for EPs and massive SEs are almost 10 times
lower than those obtained with the nominal model.  SEs with
intermediate mass ($M_p\sim 4 M_\oplus$) are less sensitive to core
radius changes and the mass losses decrease by a factor of 2 only,
when the core radius is increased.  These results support to the
conclusion that Gl 581d and HD 85512b would be well protected by their
planetary dynamos (if they actually exist or existed in the past) but
that GJ 667Cc has been too exposed to the erosive action of the
stellar wind and probably has lost its atmosphere or its content of
critical volatiles.
%REVISED2
 
In our conservative model we neglected the effect that an enhanced
early stellar magnetic activity would have in the early erosion of the
magnetically protected atmospheres, especially in the case of close-in
planets around dM stars.  It is interesting to consider how our
results could be affected if a planet were exposed to more stressful
conditions, e.g. coronal mass ejections and/or higher levels of XUV
irradiation.  We have performed sensitivity tests to study the effects
of an enhanced stellar wind and larger fluxes of XUV radiation such as
that expected for very active stars.  To achieve this we modify the
stellar wind pressure by maintaining the nominal velocity of the
plasma (as predicted by the Parker's model) but increasing by a factor
of 2 the density of stellar wind plasma\footnote{Under typical
  conditions of solar CME, the velocity of the wind is not modified
  too much but the plasma densities scales up to 5-6 times the average
  particle density}.  For the XUV flux we multiply by 2 the nominal
XUV luminosity of G-K stars (Earth-mass planet) and maintain the
initial high value of the XUV luminosity of dM stars (4.5 and 6.0
$M_\oplus$ cases) simulating an extended period of stellar activity.
The results of these sensitivity tests are depicted in the CME and
$F_\rom{XUV}$ columns in figure \ref{fig:SensitivityTest}.
%REVISED2

CME conditions change the standoff radius is only 10\%.  Again this
could be explained by the weak dependence of this quantity on the
stellar wind pressure and in particular to the number density of wind
particles, eq. (\ref{eq:Rs}) $R_S\sim P_\rom{sw}^{-1/6}\sim n^{-1/6}$.
A sustained increase in the XUV fluxes does not modify the standoff
distance at all but it is able to noticeably increase the exobase
radius and hence the time of exposure to the stellar wind erosion.  As
expected, the planet more susceptible to this effect is the Earth-mass
planet which has a N/O-rich atmosphere.  Even under harsh conditions
of stellar aggression the mass-losses for TPs, irrespective of their
mass, are no larger than 10 times the nominal value.  Again this
reinforces the conclusion that Gl 581d and HD 85512b are well
protected by their potential PMFs.
%REVISED2

The existence of mobile lids is also a key factor for the existence of
strong enough PMFs on habitable planets and it is required for the
validity of the thermal evolution model results used in this work.
The problem is being explored from different perspectives and using
complimentary methodologies (see e.g. \citealt{Valencia07c} and
\citealt{Oneill07}.  A consensus about the possibility of planets with
greater mass having or not having lid activity has not yet been
reached.  However, different lines of evidence point to the fact that,
in a wide range of rheological and thermal paramters, mobile lids
could be common in planets up to $10 \ME$ (for recent results see
e.g. \citealt{Noack12}).
%REVISED2

To estimate the radius of the exobase our model relies on the results
of detailed hydrodynamical and thermodinamic models of highly
irradiated atmospheres of TPs.  However, we have used results
calculated for two different types of atmosphere.  Although in nature
atmospheric composition could also depend on planetary mass, for
reliable comparisons it is better to use a single model.  We also have
a ``void'' in model results for planets in the mass range $1-6 \ME$.
Although we have filled the $3.5-6$ $\ME$ mass range using a simple
linear extrapolation it is expected that future improvements to the
atmospheric models will provide reliable values for the exosphere
properties for planets in that interval of masses.  We are confident
however that the exobase radius will not be too different from that
used in this work and the main qualitative conclusions will not be
modified significantly.
%REVISED2

%%%%%%%%%%%%%%%%%%%%%%%%%%%%%%%%%%%%%%%%%%%%%%%%%%%%%%%%%%%%%%%%%%%%%%%%%
\section{Summary and Conclusions}
\label{sec:Conclusions}
%%%%%%%%%%%%%%%%%%%%%%%%%%%%%%%%%%%%%%%%%%%%%%%%%%%%%%%%%%%%%%%%%%%%%%%%%

In the last few years we have seen significant improvements in the
understanding of the role that planetary magnetic fields could have in
the stability of the atmopsheres of exoplanets.  In line with these
advances in this paper we have developed a comprehensive model of the
evolution of magnetic protection of potentially habitable TPs,
integrating in a single framework the results from very different
specific areas of research in this field: thermal evolution of solid
planets, scaling of dynamo-generated magnetic fields, magnetosphere
modeling, physical properties of low-mass stars, stellar wind
evolution and atmospheric modeling of highly irradiated planets.
%REVISED2

Using this model we have studied the magnetic protection of
hypothetical habitable TPs in a wide range of planetary masses and
have addressed for the first time the cases of the already discovered
TPs found in the HZ of their host stars.  In all these cases we have
estimated the evolution of two key properties: the magnetopshere size
as measured by the standoff distance and the radius of the exobase.
The direct comparison between these properties gives us information
about the evolution of the protection that the magnetosphere provides
to the planetary atmosphere against the erosive action of the stellar
wind.  In order to estimate at to extent this magnetic protection
prevents the loss of a significant fraction of the mass of the
atmosphere or the loss of large amounts of critical volatiles
(e.g. H$_2$O and CO$_2$) we have estimated the thermal-induced
mass-loss.  Knowing that non-thermal losses could be much larger, our
conservative model provides an underestimation of the stellar wind
exposure effect.  The potentially habitable TPs that under our model
result in unsuitable conditions for magnetic protection in reality are
even worse than the model predicted.
%REVISED2

Our model is sensitive to a number of factors that affect the
quantitative results we have obtained here.  We have studied the
sensitivity of our results to the expected variations in the model
parameters and observed that the global conclusions are still very
robust.
%REVISED2

We confirm that Earth-like planets, irrespective the composition of
their atmospheres and even under the highest attainable
dynamo-generated magnetic field strengths, will lose a significant
fraction of their atmospheres or their critical volatile content if
they are tidally locked in the HZ of dM stars.  The case for the
absence of habitable Earth-like planets around this type of star are
almost closed.  The case of habitable EPs around GK stars is not as
good as previously expected either.  Earth-mass planets with N/O-rich
atmospheres, even under the best conditions of magnetic protection,
will probably lose their atmospheres or their content of water if they
are in HZ closer than $\sim$ 0.8 AU.  This limit excludes a large
range of stellar masses (0.6-0.9 $M_\odot$) depending on the
particular region inside the HZ (close to the inner or outer limit)
where the planet resides.
%REVISED2

super-Earths with $M_p\gtrsim 3 M_\oplus$ seem to have better chances
of preserving their atmospheres even if they are tidally locked around
dM stars.  Under similar conditions of thermal and magnetic evolution
there seems to exist a planetary mass-dependent inner limit inside the
HZ itself below which large atmospheric mass-losses are expected.  We
coined here the name {\it Magnetically-restricted Habitable Zone} or
MHZ for this hypothetical subregion and expect that it could be
confirmed by future improvements in the model.  This inner limit
decreases with increasing planetary mass.  Under the nominal value of
the parameters used in our conservative model we predict that for $4
\ME$ planets the limit is close to 0.15 AU, while for $8 \ME$ it will
be approximately 0.04 AU.  It implies that planets with $M_p\leqslant
4 \ME$ in HZs closer than 0.15 AU will be too exposed and probably
lose their habitable conditions in the first few Myr to 1 Gyr.  This
is precisely the case of Gj 667Cc that we predict here although inside
the HZ of its host star it is currently uninhabitable given the early
loss of its atmospheric content.  Very massive SEs ($M_p>8 \ME$) will
not have, under our conservative estimates, any restrictions and could
preserve their atmospheres even if they are in the HZ of the lowest
mass dM stars.
%REVISED2

The already discovered potentially habitable SEs Gl 581d ($M_p=6.0
M_\oplus$, $a=0.22$ AU) and HD 85512b ($M_p=3.5 M_\oplus$, $a=0.26$
AU), if assumed similar in composition to Earth, are well inside the
MHZ for their respective masses.  Even if they are subjected to larger
levels of stellar aggression their atmospheres seem to have been safe
against the strong early erosion and probably still are there.
%REVISED2

% %%%%%%%%%%%%%%%%%%%%%%%%%%%%%%%%%%%%%%%%%%%%%%%%%%%%%%%%%%%%%%%%%%%%%%
% \section*{Acknowledgments}
% %%%%%%%%%%%%%%%%%%%%%%%%%%%%%%%%%%%%%%%%%%%%%%%%%%%%%%%%%%%%%%%%%%%%%%

% \section*{Acknowledgments}

\acknowledgments

%% We want to thank to U.R. Christensen who kindly provide us detailed
%% results of numerical dynamo experiments.  We appreciate the useful
%% discussion and comments of Mercedes Lopez.  We are also grateful
%% with Luz Angela Cubides for the final revision of the manuscript
%% and their useful hints to improve the style of the text.  
%% AKNOWLEDGE: Española for useful comments and discussions.
%% on the content and the style of the manuscript which finally led to
%% its final form. 

We appreciate the useful discussion and comments of Mercedes Lopez and
other colleagues participating in the Strange New World 2011
conference (Arizona, U.S.)  and in the {\it Taller de Ciencias
  Planetarias} 2012 (Montevideo, Uruguay).  We want to give special
thanks to all our fellow colleagues abroad that have provided us with
some key literature unobtainable from our country.  We are grateful to
Peter Browning for his careful revision of the English in the first
version of the manuscript.
%% Finally we thank the anonymous referees who made many useful
%% comments on the content and the style of the manuscript which
%% finally led to its final form.
PC is supported by the Vicerrectoria de Docencia of the University of
Antioquia.  This work has been completed with the financial support of
the CODI-UdeA under the project IN591CE and under the project number
530 funded by the University of Medellin.
%REVISED2

%%%%%%%%%%%%%%%%%%%%%%%%%%%%%%%%%%%%%%%%%%%%%%%%%%%%%%%%%%%%%%%%%%%%%%%%%%%%%%%%%
%BIBLIOGRAPHY
%%%%%%%%%%%%%%%%%%%%%%%%%%%%%%%%%%%%%%%%%%%%%%%%%%%%%%%%%%%%%%%%%%%%%%%%%%%%%%%%%

%% \bibliography{bibliography}
%% \bibliographystyle{apj2}
%% % \bibliographystyle{apsrmp4-1}

\end{document}